\documentclass[sigconf, nonacm]{acmart}

\newcommand\vldbdoi{XX.XX/XXX.XX}
\newcommand\vldbpages{XXX-XXX}
\newcommand\vldbvolume{17}
\newcommand\vldbissue{X}
\newcommand\vldbyear{2024}
\newcommand\vldbauthors{Anh L.~Mai, Pengyu Wang, Azza Abouzied, Matteo Brucato, Peter J.~Haas, and Alexandra Meliou}
\newcommand\vldbtitle{\shorttitle} 
\newcommand\vldbavailabilityurl{https://github.com/alm818/PackageQuery}
\newcommand\vldbpagestyle{plain}

\usepackage{longtable}
\usepackage{tabularx}
\usepackage{xspace}
\usepackage{xcolor}
\usepackage{enumitem}
\usepackage{soul}

\usepackage{multirow}
\usepackage{colortbl}

\usepackage{amsmath}

\usepackage{booktabs}
\usepackage{graphicx}
\usepackage{subcaption}

\usepackage{algorithm}
\usepackage[noEnd=true, indLines=true]{algpseudocodex}
\usepackage{fancyvrb}
\usepackage{adjustbox}

\usepackage{libs/apxproof}
\renewcommand{\appendixbibliographystyle}{acm}
\renewcommand{\appendixsectionformat}[2]{Proofs for Section~#1}
\newtheorem{definition}{Definition}
\newtheoremrep{theorem}{Theorem}
\newtheoremrep{lemma}{Lemma}

\makeatletter
\newcommand{\definename}[3]{%
  \newcommand#1{%
    \if\@usedonce{#1}#2\else#2\fi
    \write\@auxout{\string\@isusedonce{\string#1}}%
    \gdef#1{%
      #3\write\@auxout{\string\@isusedtwice{\string#1}}%
      \gdef#1{#3}%
    }%
  }%
  \expandafter\newif\csname ifonce\string#1\endcsname
}
\def\@usedonce#1{TT\fi\csname ifonce\string#1\endcsname}
\def\@isusedonce#1{\global\csname once\string#1true\endcsname}
\def\@isusedtwice#1{\global\csname once\string#1false\endcsname}
\makeatother

\definename{\massph}{$^*$}{,}
\definename{\massam}{$^*$}{}
\definename{\nyuadam}{$^+$}{,}
\definename{\nyuadpw}{$^+$}{,}
\definename{\nyuadaa}{$^+$}{}
\definename{\msr}{$^\dagger$}{,}

\usepackage[textsize=footnotesize]{todonotes}

\usepackage[most]{tcolorbox}

\tcbset{
    experimentbox/.style={
        colback=white, 
        colframe=black!0, 
        arc=0mm, 
        coltitle=black, 
        fonttitle=\bfseries, 
        colbacktitle=white, 
        attach boxed title to top left={xshift=5mm, yshift=-\tcboxedtitleheight/2}, 
        boxed title style={boxrule=0.5pt, colframe=black, sharp corners}, 
        before=\par\smallskip\noindent, 
        after=\par\smallskip 
    }
}

\newtcolorbox{expbox}{
enhanced,breakable,
boxrule=0pt,frame hidden,
borderline west={1pt}{0pt}{blue!75!black},
colback=blue!3!white,
sharp corners,boxsep=1pt,left=4pt,right=0pt,top=4pt,bottom=0pt
}

\newtheorem{alg}{Algorithm}[section]

\DeclareMathOperator*{\argmax}{arg\,max}

\newcommand{\ra}[1]{\textcolor{red}{#1}}

\definecolor{mygreen}{rgb}{0,0.6,0}
\definecolor{mymauve}{rgb}{0.58,0,0.82}

\newcommand{\rb}[1]{{\color{mygreen}{#1}}}
\newcommand{\rc}[1]{{\color{mymauve}{#1}}}
\newcommand{\common}[1]{\textcolor{blue}{#1}}

\renewcommand{\ra}[1]{\textnormal{#1}}
\renewcommand{\rb}[1]{\textnormal{#1}}
\renewcommand{\rc}[1]{\textnormal{#1}}
\renewcommand{\common}[1]{\textnormal{#1}}

\algrenewcommand\algorithmicensure{\textbf{Parameter:}}

\newcommand{\lsr}{\textsc{Progressive Shading}\xspace}

\newcommand{\shading}{\textsc{Shading}\xspace}

\newcommand{\sr}{\textsc{SketchRefine}\xspace}

\newcommand{\dlv}{\textsc{Dynamic Low Variance}\xspace}
\newcommand{\adlv}{\textsc{DLV}\xspace}

\newcommand{\kd}{\textsc{kd-tree}\xspace}
\newcommand{\pg}{\textsc{PostgreSQL}\xspace}

\newcommand{\filp}{\textsc{Dual Reducer}\xspace}
\newcommand{\neighbor}{\textsc{Neighbor Sampling}\xspace}

\newcommand{\getILP}{\textsc{Formulate ILP}\xspace}
\newcommand{\getLP}{\textsc{Formulate LP}\xspace}

\newcommand{\pds}{\textsc{Parallel Dual Simplex}\xspace}

\newcommand{\paql}{\textsc{PaQL}\xspace}

\newcommand{\gb}{\textsc{Gurobi}\xspace}

\newcommand{\cp}{\textsc{CPLEX}\xspace}

\newcommand{\ssds}{\textsc{SDSS}\xspace}
\newcommand{\tpch}{\textsc{TPC-H}\xspace}

\newcommand{\onedlv}{\textsc{1-D Dynamic Low Variance}\xspace}

\newcommand{\aonedlv}{\textsc{1-D DLV}\xspace}

\newcommand{\hard}{\ensuremath{\Tilde{h}}\xspace}

\newcommand{\xP}{\mathcal{P}}
\newcommand{\aug}{\hyperlink{augdef}{\textit{augmenting size}}\xspace}
\newcommand{\df}{\hyperlink{dfdef}{\textit{downscale factor}}\xspace}

\newcommand{\augv}{\hyperlink{augdef}{\ensuremath{\alpha}}\xspace}

\newcommand{\dfv}{\hyperlink{dfdef}{\ensuremath{d_f}}\xspace}

\usepackage{tikz}
\newcommand\defbox[2][fill=black!10]{%
    \tikz[baseline]\node[%
        inner ysep=0pt, 
        inner xsep=2pt, 
        anchor=text, 
        rectangle, 
        rounded corners=1mm,
        #1] {\strut#2};%
}

\newcommand{\eat}[1]{{}}

\usepackage[normalem]{ulem}

\title{Scaling Package Queries to a Billion Tuples via Hierarchical Partitioning and Customized Optimization}

\author{Anh L.~Mai$^+$\quad Pengyu Wang$^+$\quad Azza Abouzied$^+$\\ \quad Matteo Brucato$^\dagger$\quad Peter J.~Haas$^*$\quad Alexandra Meliou$^*$}

\affiliation{
\begin{tabular}{c c c}
    {$^+$}\institution{New York University Abu Dhabi}\country{}&
    {$^\dagger$}\institution{Microsoft Research}\country{}&
    {$^*$}\institution{University of Massachusetts Amherst}\country{}\\
    \institution{\{anh.mai, pengyu.wang, azza\}@nyu.edu} & 
    \institution{mbrucato@microsoft.com} &
    \institution{\{phaas, ameli\}@cs.umass.edu}
\end{tabular}
}

\begin{document}

\begin{abstract}

A package query returns a package---a multiset of tuples---that maximizes or minimizes a linear objective function subject to linear constraints, thereby enabling in-database decision support.  Prior work has established the equivalence of package queries to Integer Linear Programs (ILPs) and developed the \sr algorithm for package query processing. While this algorithm was an important first step toward supporting prescriptive analytics scalably inside a relational database, it struggles when the data size grows beyond a few hundred million tuples or when the constraints become very tight. In this paper, we present \lsr, a novel algorithm for processing package queries that can scale efficiently to billions of tuples and gracefully handle tight constraints. \lsr solves a sequence of optimization problems over a hierarchy of relations, each resulting from an ever-finer partitioning of the original tuples into homogeneous groups until the original relation is obtained. This strategy avoids the premature discarding of high-quality tuples that can occur with \sr. Our novel partitioning scheme, \dlv, can handle very large relations with multiple attributes and can dynamically adapt to both concentrated and spread-out sets of attribute values, provably outperforming traditional partitioning schemes such as \kd. We further optimize our system by replacing our off-the-shelf optimization software with customized ILP and LP solvers, called \filp and \pds respectively, that are highly accurate and orders of magnitude faster.
\end{abstract}

\maketitle

\pagestyle{\vldbpagestyle}
\begingroup\small\noindent\raggedright\textbf{PVLDB Reference Format:}\\
\vldbauthors. \vldbtitle. PVLDB, \vldbvolume(\vldbissue): \vldbpages, \vldbyear.\\
\href{https://doi.org/\vldbdoi}{doi:\vldbdoi}
\endgroup
\begingroup
\renewcommand\thefootnote{}\footnote{\noindent
This work is licensed under the Creative Commons BY-NC-ND 4.0 International License. Visit \url{https://creativecommons.org/licenses/by-nc-nd/4.0/} to view a copy of this license. For any use beyond those covered by this license, obtain permission by emailing \href{mailto:info@vldb.org}{info@vldb.org}. Copyright is held by the owner/author(s). Publication rights licensed to the VLDB Endowment. \\
\raggedright Proceedings of the VLDB Endowment, Vol. \vldbvolume, No. \vldbissue\ %
ISSN 2150-8097. \\
\href{https://doi.org/\vldbdoi}{doi:\vldbdoi} \\
}\addtocounter{footnote}{-1}\endgroup

\ifdefempty{\vldbavailabilityurl}{}{
\vspace{.3cm}
\begingroup\small\noindent\raggedright\textbf{PVLDB Artifact Availability:}\\
The source code, data, and/or other artifacts have been made available at \url{\vldbavailabilityurl}.
\endgroup
}

\section{Introduction}
\label{sec:intro}

\emph{Package queries}~\cite{brucato} extend traditional relational database queries to handle constraints that are defined over a multiset of tuples called a ``package.'' A package has to satisfy two types of constraints:
\begin{itemize}[leftmargin=*]
    \item \emph{Local predicates}: traditional selection predicates, i.e., constraints that each tuple in the package has to satisfy individually.
    \item \emph{Global predicates}: constraints that all the tuples within the package have to satisfy collectively.
\end{itemize}
There can be many such feasible packages. A package query selects a feasible package that maximizes or minimizes a linear objective.

\begin{figure*}[t]
\centering
\begin{subfigure}[h]{0.195\textwidth}
  \centering
  \includegraphics[width=\textwidth]{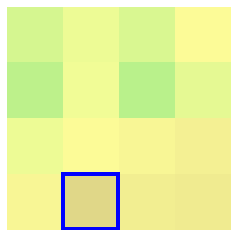}
\end{subfigure}
\begin{subfigure}[h]{0.195\textwidth}
  \centering
  \includegraphics[width=\textwidth]{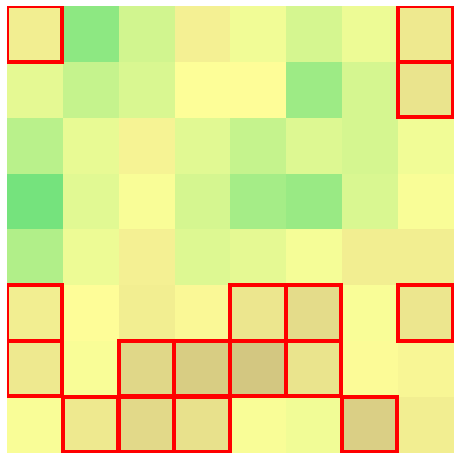}
\end{subfigure}
\begin{subfigure}[h]{0.195\textwidth}
  \centering
  \includegraphics[width=\textwidth]{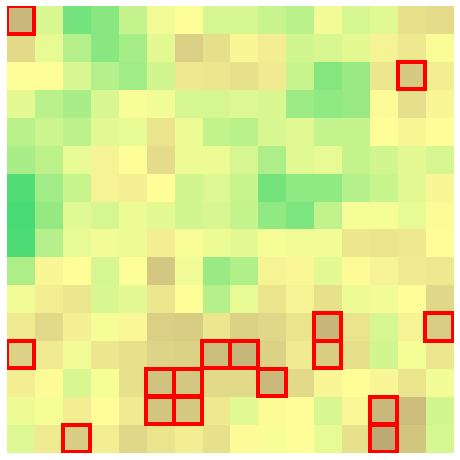}
\end{subfigure}
\begin{subfigure}[h]{0.195\textwidth}
  \centering
  \includegraphics[width=\textwidth]{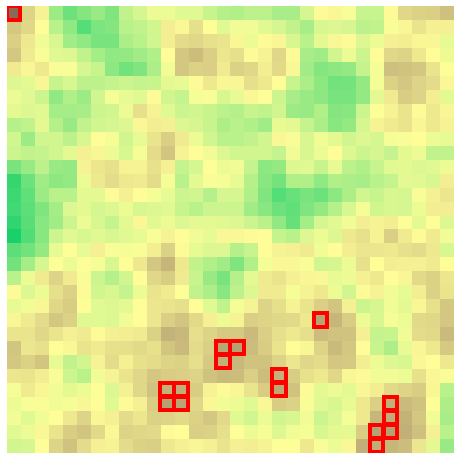}
\end{subfigure}
\begin{subfigure}[h]{0.195\textwidth}
  \centering
  \includegraphics[width=\textwidth]{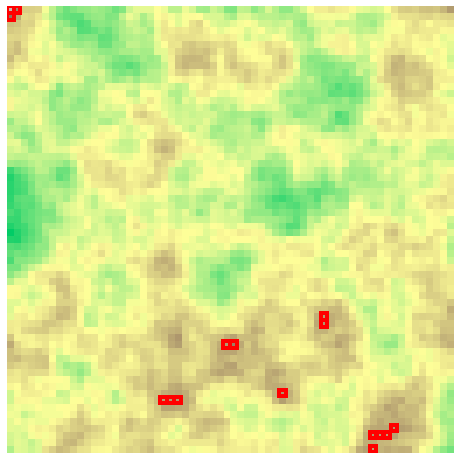}
\end{subfigure}
\vspace{-2mm}
\caption{From left to right: five increasing resolutions of terrain height: 4x4, 8x8, 16x16, 32x32, 64x64. The blue square is the highest square in the lowest resolution. The red squares are the 16 highest squares in each subsequent resolution.}
\label{fig:res}
\vspace{-2mm}
\end{figure*}

\looseness-1
For example, consider the following package query: \textit{An astrophysicist needs to find a certain number of rectangular regions of the night sky that may contain unseen quasars. These regions should have average brightness above a certain threshold and their overall red shift should lie between specified values. Among those regions, the one with the maximum combined log-likelihood of containing a quasar is preferred} \cite{window1}. \ra{Suppose that we have a \texttt{Regions} table as below:
\smallskip
\begin{center}\small
\begin{tabular}{|cccccc|c}
\multicolumn{1}{c}{\textbf{ID}} &  \textbf{brightness}& \textbf{redshift} & \textbf{quasar} & $\cdots$ & \multicolumn{1}{c}{\textbf{explored}}&\\
\cline{1-6}
301 & 6.0 & 1.47 & -0.05 & $\cdots$ & true& $x_1$\\
491 & 9.6 & 1.68 & -0.01 & $\cdots$ & false& $x_2$ \\
$\vdots$ & $\vdots$ & $\vdots$ & $\vdots$ & $\vdots$ & $\vdots$ &$\vdots$\\
\cline{1-6}
\end{tabular}
\end{center}}
 \smallskip
\noindent This package query can be expressed declaratively using \paql, an SQL-based query language \cite{brucato}: 

\begin{Verbatim}[commandchars=\\\{\},codes={\catcode`$=3\catcode`_=8}]
SELECT      \textbf{PACKAGE}(*) AS P
FROM        Regions R \textbf{REPEAT} 0
WHERE       R.explored = 'false'
\textbf{SUCH THAT}   COUNT(P.*) = 10
            AVG(P.brightness) $\geq \theta$
            SUM(P.redshift) BETWEEN $\gamma_1$ AND $\gamma_2$
\textbf{MAXIMIZE}    SUM(P.quasar)
\end{Verbatim}

\noindent In this example, the number of rows, i.e., rectangular regions of the night sky, can become very large if the surveying resolution is high and/or the surveying volume of the night sky is large. For such a query, the relation size typically ranges from millions to billions of regions while the number of constraints is constant. \rb{This example shows how a package query with a very large number of rows can arise in scientific applications such as astronomy, oceanography, atmospheric science, and more. Large package queries can appear in many other domains, in the context of decision support. For example, consider a national marketing campaign where each person is exposed to one out of a possible set of $k$ personalized ads. Each row of the table now corresponds to a (person, ad) pair. A model predicts the expected purchase amount for each person, given the person's features and the ad. The goal is to select an ad for each person so as to maximize predicted sales, subject to constraints on the advertising budget. The problem becomes even larger if each person can be shown multiple ads over a period of time. Other examples include certain types of portfolio optimization problems~\cite{kuhn2023integer}}.

\looseness-1
Every package query corresponds to an Integer Linear Program (ILP)~\cite{brucato}, a common but challenging type of optimization problem. For a relation containing $n$ tuples, there are $n$ decision variables, with the $i$th decision variable $x_i$ representing the multiplicity (possibly 0) of the $i$th tuple in the package. \ra{In the astrophysics example, $x_i=1$ if the $i$th region is included in the package and $x_i=0$ otherwise. Thus, setting $c_i=t_i.\text{quasar}$,  we want to maximize the linear function  $\sum_i c_ix_i$ subject to linear constraints such as $\sum_i x_i=10$ and $\sum_i a_ix_i\le \gamma_2$. where $a_i=t_i.\text{redshift}$.} Thus, black-box ILP solvers like \gb~\cite{gurobi} or \cp~\cite{cplex} can, in principle, be used to compute the optimal package for any package query. When $n$ grows beyond several million, however, the foregoing solvers typically do not scale because they employ ILP techniques that have  $O\bigl(\exp(n)\bigr)$ worst-case running time. The \sr approximate package-query processing algorithm introduced in \cite{brucato} breaks down the original optimization problem into a sequence of small problems and works well up to tens of millions of decision variables. Beyond this scale, however, its performance deteriorates in both running time and optimality as shown by our experimental results in Section~\ref{sec:results}. \ra{Because the number of decision variables in a package query is often multiple orders of magnitude greater than the ``large'' problems previously studied in the optimization literature, prior approximate ILP algorithms have even more trouble scaling because, unlike \sr, they need to process all the decision variables at once~\cite{milp_heuristic}.}

\vspace{2pt}
\noindent
\textbf{Overview of \sr and its limitations.} \sr partitions relations to scalably approximate package queries. Each partition contains similar tuples that are averaged to construct a representative tuple. In \sr, a ``sketch'' is a package solution over the representative tuples only. Representative tuples included in the sketch indicate that their groups may have tuples that can be part of the optimal package. The sketch is ``refined'' by searching through these groups, iteratively replacing each representative with the group's tuples, and re-solving the package-query ILP until a feasible package is constructed from the actual tuples.

\looseness-1
\sr uses \kd partitioning \cite{kdtree} with a fixed number of groups, regardless of the relation size. The number of groups is usually small (e.g., up to 1000 groups for a relation size of tens of millions) resulting in a large number of tuples in each group. While this approach allows aggressive pruning of the relation in the sketch phase, it has three drawbacks. First, the representative tuples may not accurately represent their groups, especially if the underlying distribution of tuples has a high variance. This can lead to \emph{false infeasibility}, where no solution is found during sketching even though a feasible package does exist. Our experimental results in Section \ref{sec:results} show that the prevalence of false infeasibility increases significantly as the query constraints become tighter, i.e., as the feasible region shrinks. Second, the aggressive pruning of entire groups might eliminate potential tuples at the periphery of these groups from consideration, i.e., groups that were not selected in the sketch can contain outlying tuples that can improve the overall objective value. Without these tuples, \sr can produce packages with suboptimal objective values. Third, when the relation size increases, the size of the refine queries, which essentially equals the group size, increases. It is challenging at best, and often impossible in practice, to decide exactly how fine the partitions should be. Creating too many groups will result in high computational costs for the partitioning algorithm and for solving the sketch query. Creating too few groups will degrade the accuracy of the sketch query and possibly cause false-infeasibility problems, as well as rendering the solution of each refine query hugely expensive. \sr thus fails to scale to relations on the order of 100M tuples or more.

\vspace{2pt}
\noindent
\textbf{Our new approach.}
In this work, we introduce \lsr, a novel approach for approximately solving package queries over extremely large relations which overcomes the above limitations. Our method relies on a hierarchy of relations comprising $L+1$ layers of increasingly aggregated representative tuples. Layer~$0$ consists of the original tuples from the relation; each layer~$l>0$ comprises representative tuples (representing groups) obtained after partitioning the tuples in layer~$l-1$. Each layer comprises a large number of groups---the size of each group is small so that each representative tuple in layer~$l$ accurately summarizes the attribute values of its corresponding tuples in layer~$l-1$. The search for optimal packages starts in layer~$L$ by solving a linear program (LP) over all of its representative tuples under the original constraints and objective but with the integrality requirement on the decision variables removed. The chosen tuples found in the LP are augmented with additional nearby "promising" representative tuples to help prevent premature discarding of potentially valuable tuples in layer~$L-1$. Then all of the chosen tuples in layer~$L$ are expanded into their corresponding groups in layer~$L-1$. This so-called \neighbor procedure of augmenting and expanding representative tuples from the current LP solution is executed at each successive level of the hierarchy until we reach layer~$0$, at which point we solve a final ILP to produce the solution package (Section \ref{sec:neighborSearch}). 

Intuitively, the differences between \sr and the iterative procedure of \lsr can be described via an analogy to resolution-mapping techniques in fields like geographic or demographic analysis~\cite{reso}. Figure \ref{fig:res} shows a hierarchy of resolution from low to high of a terrain-height map that is analogous to the hierarchy of relations in \lsr. An efficient approach like \lsr would start from the lowest resolution and iterate toward the highest resolution. In each iteration, it starts with the 16 highest squares in the current resolution and expands those squares into the next resolution, and, among those, selects the 16 highest squares. This approach diversifies the final result in the highest resolution and thus, captures the irregularities of the terrain. On the other hand, \sr is analogous to simply looking at the highest square (the blue square) in the lowest resolution and analyzing its height in the highest resolution.

The former resolution-mapping approach only works when each resolution is not downscaled too drastically to the next lower resolution---otherwise, the search within each higher-resolution square takes too long. For example, the downscale factor from 64x64 to 4x4 resolution is 256 since 5376 pixels are partitioned into 16 squares. On the other hand, going from 64x64 to 32x32 resolution has a downscale factor of 4. Analogously, the hierarchy of relations in \lsr requires a partitioning algorithm that: 
\begin{itemize}[topsep=1pt,leftmargin=*]
    \item Efficiently produces a large number of partitions/groups, e.g., typically about 0.1\%-10\% of the number of tuples (so a downscale factor between 1000 and 10). A typical \kd partition used in \sr \cite{brucato} has a downscale factor of $n/g$ where $n$ is the relation size and $g$ is the fixed number of groups (at most 1000), which explains why \kd is not particularly suitable for \lsr when $n$ is large; and
    \item Supports fast group-membership determination for arbitrary tuple values (not necessarily appearing in the relation), as needed for efficient execution of \lsr.
\end{itemize}

\noindent
We, therefore, provide a novel partitioning algorithm, \dlv (\adlv), that satisfies these requirements. The advantages of \adlv over standard partitioning algorithms such as $k$-means~\cite{kmean}, hierarchical clustering~\cite{hiearchy}, and $k$-dimensional quad-trees~\cite{kdtree} are (1)~its ability to run under limited memory, (2)~its cache-friendliness, and (3)~its high parallelizability. Importantly, \adlv is a \textit{dynamic} scheme, which allows it to refine its partitions in response to outliers, i.e., to the shape of the distribution of the tuple's attributes: in our stylized example, a \adlv partition on the 64x64 resolution can isolate high peaks into their own groups to maintain low variance within groups. \adlv minimizes attribute variance to implicitly ensure that similar tuples are grouped together.

At the end of \lsr, we end up with an in-memory ILP of a package query with tuples from the original relation. This ILP typically has at least hundreds of thousands of variables. Black-box ILP solvers would require a large amount of time to produce an optimal solution (especially when the underlying ILP is hard to solve) and hence, are unsuitable for \lsr. We, therefore, develop \filp, a new heuristic algorithm that can solve a package query over millions of tuples in less than a second with close-to-optimal objective values. It achieves this by first solving an LP that essentially removes the integrality constraints of the ILP and then formulates a second LP using constraints that help prune tuples whose corresponding decision variables likely will not appear in the ILP solution. It effectively shrinks the original ILP into a very small sub-ILP that can be efficiently handled by black-box ILP solvers. As with any heuristic ILP solver, false infeasibility can occur in \filp when the pruning is too aggressive. We handle this issue by gradually reducing the degree of pruning until we end up solving the original ILP using a black-box ILP solver.

\begin{figure}[t]
    \centering
    \includegraphics[width=\linewidth]{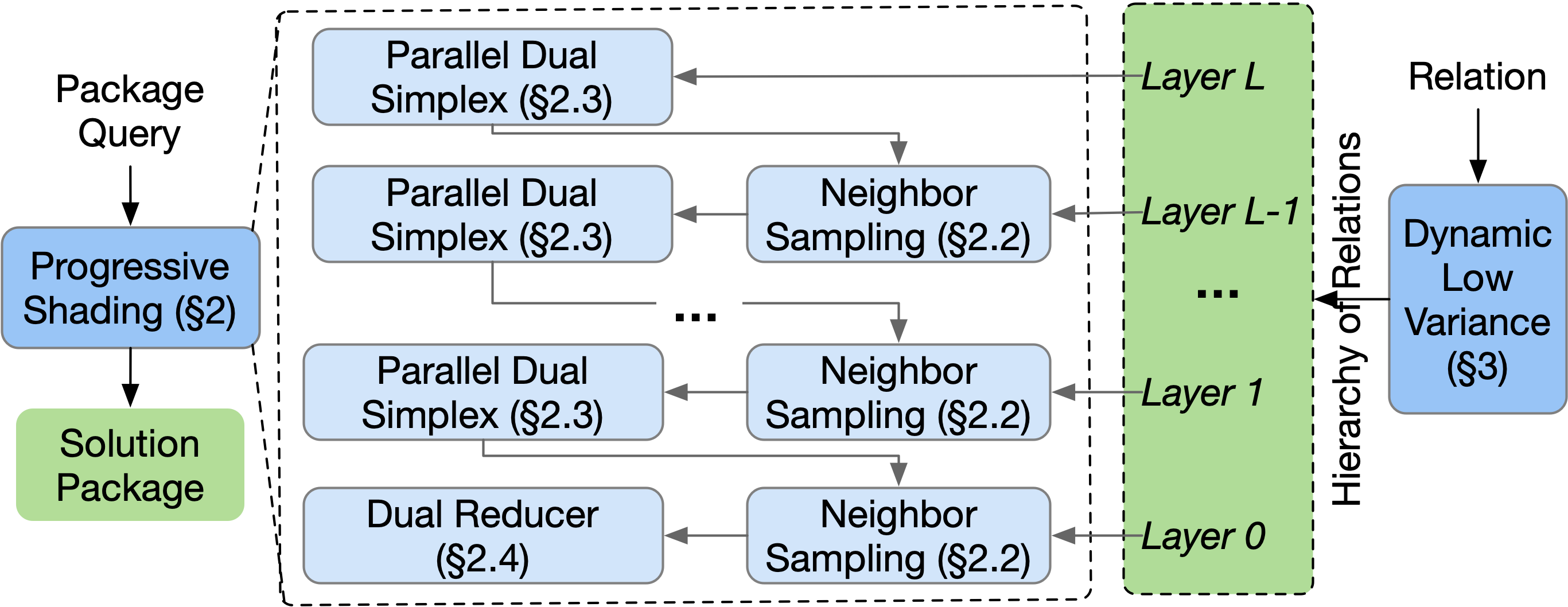}
        \vspace{-6mm}
    \caption{High-level architecture of \lsr and \adlv for scaling package query evaluation over very large relations.} 
    \label{fig:summary}
            \vspace{-5mm}
\end{figure}

\lsr extensively uses an LP solver for its intermediate layers and an ILP solver for layer~0. \ra{To further boost performance, we replace the intermediate black-box LP solver with our highly accurate and much faster implementation \pds, which exploits the special structure of the ILPs that arise when solving package queries compared to general ILPs (Section~\ref{sec:pds}). We also replace the final black-box ILP solver with our novel \filp heuristic ILP solver (Section~\ref{sec:filp}); see Figure~\ref{fig:summary}.}

\smallskip\noindent\textbf{Contributions.} In summary, we significantly expand the applicability of package-query technology to handle very large problems with potentially tight constraints via the following contributions:
\begin{itemize}[leftmargin=*]
    \item A novel hierarchical strategy, called \lsr, for finding high-quality package tuples that avoids the pitfalls of the \sr approach (Section~\ref{sec:lsr}).

    \item An effective and efficient partitioning scheme, \dlv, for creating the hierarchical data partitions needed by \lsr and handling outlying data well, together with an analytical comparison to \kd that verifies \adlv's superior behavior (Section~\ref{sec:dlv}).

    \item A novel heuristic, \filp, for a very fast approximate solution of the final ILP encountered in \lsr that uses a simple pruning strategy and a mechanism to guarantee solvability (Section~\ref{sec:filp}), along with an optimized and highly parallelized LP solver, \pds, for accurate solution of LPs encountered in \lsr (Section~\ref{sec:pds}). 
   
    \item A thorough experimental study showing that, unlike \sr, \lsr is scalable beyond hundreds of millions of tuples and, even for smaller relations, it can solve ``hard'' package queries for which \sr suffers from false infeasibility. When both algorithms can produce feasible packages, \lsr is faster and the solution packages have better objective values (Section~\ref{sec:results}). Notably, as part of this study, we define a novel  \textit{hardness} metric and provide a way to generate queries of a specific hardness, thereby providing a means and a benchmark to systematically evaluate package-query solvers (Section~\ref{sec:Esetup}).
\end{itemize}
\section{\lsr}
\label{sec:lsr}

The key challenge with directly solving the large ILPs that arise from package queries over large relations is that current solvers require that the corresponding LPs (where the integrality constraints are removed) fit into memory. Both \sr and \lsr algorithms avoid this problem by partitioning the large relation into smaller groups that fit in memory and then formulating small ILPs based on the representative tuples corresponding to these groups, thereby obtaining an approximate solution to the original ILP. These two algorithms, however, use very different strategies to obtain these small ILPs.

While \sr ``refines'' the sketch solution by iteratively replacing each chosen representative tuple with the group's tuples, \lsr first \textit{augments} the sketch solution with additional ``promising'' representative tuples and then replaces all the chosen representative tuples with their group's tuples at once. In doing so, \lsr tries to make each intermediate LP as large as possible by augmentation (via \neighbor); this improves the quality of the ILP solution of the original tuples relative to \sr by not eliminating potentially high-quality tuples from consideration too early. More specifically, the algorithm tries to always solve an LP or ILP where the number of variables is close to, but does not exceed, an  upper bound \augv. We call \augv the \defbox{\hypertarget{augdef}{\textit{augmenting size}}\xspace}; it is chosen so that an LP with \augv variables fits in memory and can be solved relatively fast, e.g., within 1 second for interactive performance~\cite{tti}.

\begin{figure*}[tb]
    \centering
    \includegraphics[width=1\textwidth]{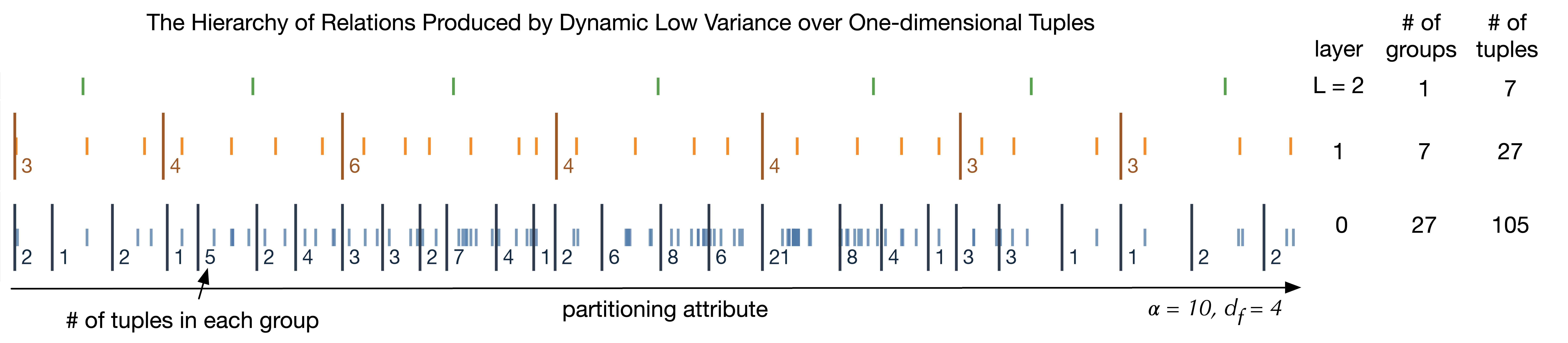}
    \caption{A 3-layer hierarchy of relations produced by \dlv with a \protect\df of 4. Each short colored bar represents a tuple in the hierarchy with long vertical black lines denoting partition boundaries.} 
    \label{fig:dlv-hierarchy}
    \vspace{-2mm}
\end{figure*}

\smallskip\noindent\textbf{Hierarchy of Relations.}
\lsr relies on a hierarchy of relations of $L+1$ layers where layer $0$ is the original relation and each layer~$l \geq 1$ is the relation comprised of $r_l$ representative tuples obtained after grouping the $n_{l-1}$ tuples in layer~$l-1$. That is, layer~$l-1$ is partitioned into $r_l$ groups with \df $\dfv = n_{l-1}/r_l$. So the \defbox{\hypertarget{dfdef}{\textit{downscale factor}}\xspace} \dfv is the average number of tuples per group.

Given \dfv, the depth~$L$ of the hierarchy is the smallest number of layers such that the final layer~$L$ has a size at most \augv. That is, for a relation having $n$ tuples and a \df \dfv, the final layer~$L$ has a size approximately $n/({\dfv})^L \leq \augv$, so that the minimal number of layers is $L=\lceil \log_{\dfv}(n/\augv)\rceil$. See Figure \ref{fig:dlv-hierarchy} for an example.

A group in layer $l\in[0..L]$ is defined by intervals $[a_j,b_j]$ where $-\infty \leq a_j < b_j \leq \infty$ for each attribute~$j$ such that all groups are non-overlapping. A tuple $t$ belongs to the group if and only if $t.j \in [a_j, b_j]$ for all $j$, where $t.j$ is the attribute~$j$ of tuple~$t$. 

To compute the hierarchy of relations, we apply our partitioning algorithm, \dlv (Section \ref{sec:dlv_multi}), iteratively from layer~$0$ to layer~$L-1$ with a \df \dfv. \rb{For ease of presentation, we largely ignore the effects of local predicates on the solvability and optimality of the solution package; see Appendix~\ref{appendix:local} for a brief discussion of how to mitigate the decreasing accuracy of representative tuples as local predicates select fewer tuples.}

\begin{algorithm}[t]
\caption{\lsr}\label{alg:lsr}
\begin{algorithmic}[1]
\Require $Q :=$ package query
\Ensure $\alpha :=$ augmenting size
\State $S_L \gets $ set of indices of all representative tuples at layer $L$
\State $l \gets L$
\While{$l > 0$}
\State $S_{l-1} \gets$ \textsc{Shading}($l, \alpha, S_l, Q$)
\State $l \gets l - 1$
\EndWhile

\State $S^* \gets$ \filp($Q$, $S_0$) 

\Return $S^*$
\end{algorithmic}
\end{algorithm}

\smallskip\noindent\textbf{\lsr overview.} A high-level view of \lsr is presented in Algorithm~\ref{alg:lsr}. Given the hierarchy of relations, along with the \aug \augv, \lsr processes a package query by
starting with the set $S_L$ of all potential candidates---i.e., the set of all representative tuples---in layer~$L$ (line~1) and then iterates through the hierarchy down to layer~$0$ using \shading (Algorithm~\ref{alg:shading}) to return a set $S_{l-1}$ of at most \augv potential candidates from layer~$l-1$ given the set of potential candidates $S_l$ from layer~$l$ (line~4). At layer~0, \lsr produces the  final solution package from the package query $Q[S_0]$ using \filp (Section \ref{sec:filp}), our heuristic ILP solver specifically designed to be efficient when solving ILPs arising from package queries (\rb{line~6}). We describe the various components of \lsr in the following subsections.

\subsection{\shading}
\label{sec:shad}

\begin{algorithm}[t]
\caption{\textsc{Shading}}\label{alg:shading}
\begin{algorithmic}[1]
\Require $l := $ layer~$l > 0$
\Statex $\augv :=$ \aug
\Statex $S_l := $ set of indices of potential candidates at layer $l$
\Statex $Q := $ package query

\State $P \gets$ \getLP($Q[S_l]$) 

\State $x^* \gets$ \pds($P$)

\State $S_l' \gets \{i \in S_l | x^*_i > 0\}$ \Comment{{\color{gray}\textit{$S_l' \subseteq S_l$}}}

\State $S_{l-1} \gets \neighbor(l, \alpha, S_l')$

\Return $S_{l-1}$

\end{algorithmic}
\end{algorithm}

\shading starts by formulating a package query $Q[S_l]$ from the tuples in $S_l$, which leads to an ILP. The algorithm then formulates an LP by removing the integrality conditions of the ILP (line 1). 

It then solves the LP using \pds (Section~\ref{sec:pds}) --- our efficient LP solver specifically designed to exploit the fact that package queries have a very low number of constraints $m$ (line~2). The LP solution $x^*$ serves only to seed the initial set $S'_l$ of potential candidates, i.e., $S'_l$ comprises tuples with positive coefficients in $x^*$ (line 3). The final step is to augment and expand the representative tuples in $S'_l$ to $S_{l-1}$ (line 4) via the \neighbor algorithm (Section \ref{sec:neighborSearch}). 

A potential concern is that expanding the representative tuples in $S'_l$ might generate an excessive number of candidate tuples at layer $l-1$; that is, the expected number of layer-$(l-1)$ tuples $\dfv |S'_l|$ will exceed \augv and removal of tuples, rather than augmentation up to size \augv, will be required.
This scenario is unlikely, though, because (1)~\dfv is typically small (Section~\ref{sec:dlv_one}), and
(2)~for package queries, the number $|S'_l|$ of positive coefficients in $x^*$ is typically small in that $|S'_l| \leq \lceil m+\lVert x^* \rVert_1 \rceil \ll \augv$ where $m$ is the number of constraints and $\lVert \cdot \rVert_1$ is the L1 norm (see Section~\ref{sec:filp} for a proof). \rb{If this scenario occurs, we can remove tuples in order of worst objective-value coefficient first until the number of layer-$(l-1)$ tuples is at most $\alpha$.}

\begin{expbox}
\textbf{Mini-Experiment 1.} \common{\textit{Does replacing the LP solution with an ILP solution in \shading improve overall optimality?}}

\common{No. } We observed no improvement in \lsr's optimality or solvability when replacing the LP solution (line~2 of the \shading algorithm) with an ILP one. As LPs are faster to solve than ILPs, we prefer the LP formulation. \common{See \cite[Figure~\ref{fig:m1}]{pq} for details.}
\end{expbox}

\subsection{\neighbor}
\label{sec:neighborSearch}

\begin{figure}[tb]
    \centering
    \includegraphics[width=0.5\textwidth]{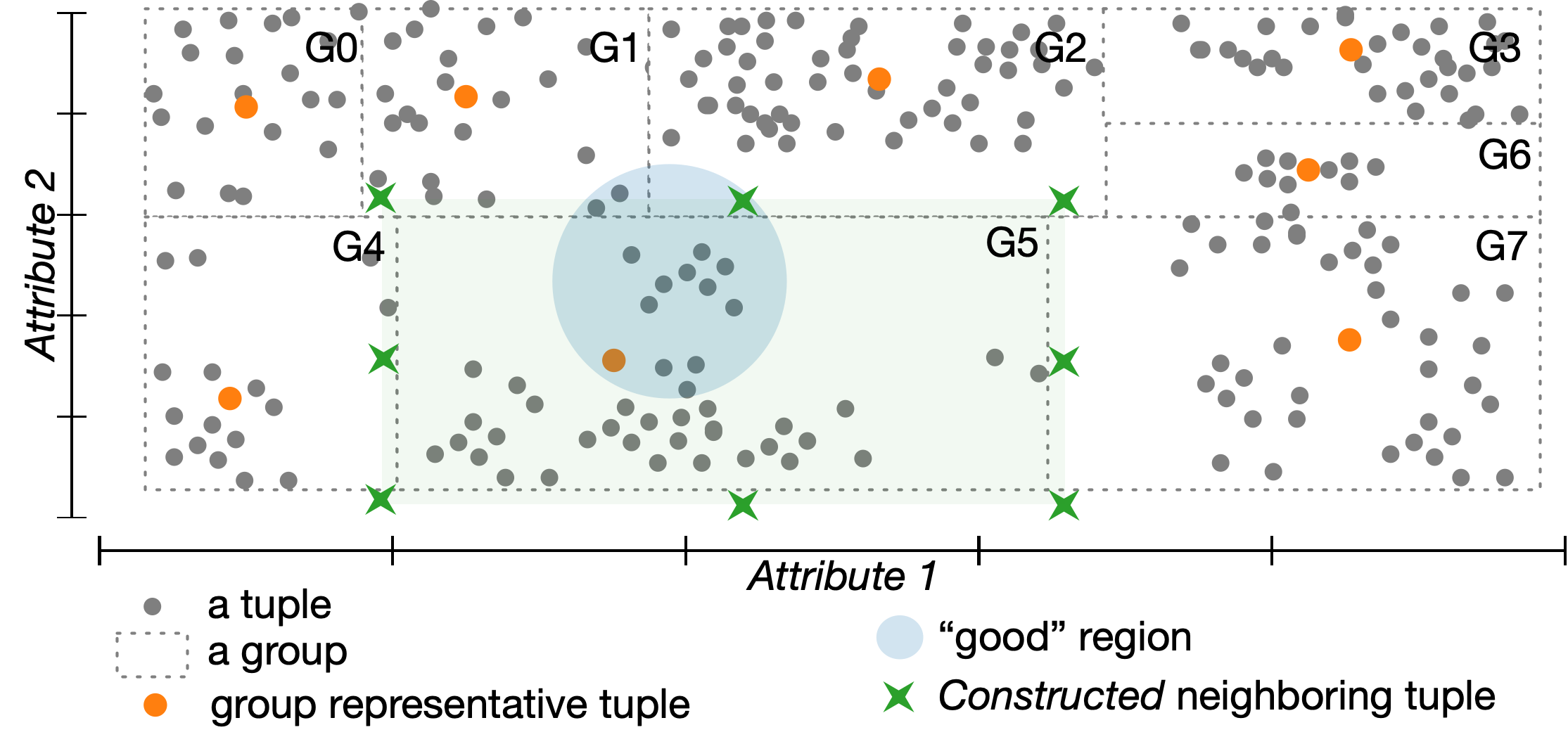}
    \caption{Groups G1, G2, G4, and G7 are neighboring to G5 since they contain the constructed neighboring tuples.} 
    \label{fig:shade}
    \vspace{-2mm}
\end{figure}

Given the solution tuples $S'_l$ at layer $l$, \neighbor in line~4 of Algorithm~\ref{alg:shading} selects tuples $S_{l-1}$ from layer $l-1$. \rb{First, it replaces $l$-layer tuple $g$ in $S'_l$ with the $l-1$-layer tuples of the group $g$ via \textsf{GetTuples}$(l-1,g)$ (line 2 of Algorithm~\ref{alg:neighbor_group}).} It then augments this set with tuples from neighboring groups.

Figure \ref{fig:shade} shows a typical representation of 2D groups, demarcated by horizontal and vertical lines. In general, a group can have more than two attributes, with $[a_j,b_j]$ specifying the group boundaries along attribute $j$. Each group is represented by its average tuple (the orange dot in Figure \ref{fig:shade}). Suppose the blue circle represents a "good" region of tuples that are likely to be found in the optimal package. G5's representative tuple (orange dot) lies within this good region and is selected in the candidate solution set $S'_l$. If we only select G5's tuples for the next \shading iteration, we would miss out on tuples in G1 that lie within the good region. These are \textit{hidden outliers} --- tuples that are potentially in the final solution but are hidden as their groups' representative tuples are far from the ``good'' region. We want an algorithm that can add tuples from these neighboring groups which are identified by some measure of ``closeness'' to the selected group G5.

\begin{algorithm}[tb]
\caption{\neighbor}\label{alg:neighbor_group}
\begin{algorithmic}[1]

\Require $l := $ layer~$l > 0$
\Statex $\alpha :=$ augmenting size
\Statex $S_l' := $ set of indices of tuples selected by the LP solution.

\State $\epsilon \gets \min_{t.j \neq \Tilde{t}.j} |t.j-\Tilde{t}.j|$

\State $S_{l-1} \gets \cup_{g \in S_l'}$ \rb{\textsf{GetTuples}$(l-1, g)$}

\State $\bar{S}_l' \gets \emptyset$ \Comment{ {\color{gray}\textit{The complement of $S'_l$}}}
\While{$|S_l'| > 0$ and $|S_{l-1}| < \alpha$}
    \State \rb{$g \gets \argmax_{g \in S'_l} \textsf{ObjVal}(g)$}
\State $S_l' \gets S_l' \setminus \{g\}$ \Comment{ {\color{gray}\textit{$S_l'$ is a max-priority queue}}}
\State $\bar{S}_l' \gets \bar{S}_l' \cup \{g\}$
\State $A_g \gets \{[a_j, b_j], j=1,...,k\}$
\State $T \gets \{a_1-\epsilon, \frac{a_1+b_1}{2}, b_1+\epsilon\}\times ... \times \{a_k-\epsilon, \frac{a_k+b_k}{2}, b_k+\epsilon\}$
\For{each $t \in T$}
    \State $g' \gets$ \textsf{GetGroup}($l,t$)
    \If{$g' \notin S_l' \cup \bar{S}_l'$} 
    \Comment{ {\color{gray}\textit{If we have not seen $g'$ before}}}
        \State $S_l' \gets S_l' \cup \{g'\}$
        \State $S_{l-1} \gets S_{l-1} \cup $ \rb{\textsf{GetTuples}$(l-1, g')$}
    \EndIf 
\EndFor
\EndWhile

\Return $S_{l-1}[:\alpha]$ \Comment{{\rb{\textit{Return $\alpha$ highest objective tuples.
}}}}

\end{algorithmic}
\end{algorithm}
 
\rb{Without loss of generality, we present the \neighbor algorithm~\ref{alg:neighbor_group} assuming an objective maximization query. One can replace `$\max$/highest' with `$\min$/lowest' for objective minimization queries.}
We select a group $g$ \rb{with the highest objective value $\textsf{ObjVal(g)}$ (line~5)}. Group $g$ is defined by a set of intervals $A_g=\{[a_j, b_j], j=1,...,k\}$ (line 8). We \textit{construct} a neighboring tuple~$t$ that lies ``just outside'' group $g$
by setting each attribute~$t.j$ equal to $a_j-\epsilon$, $b_j+\epsilon$, or $(a_j+b_j)/2$ where $\epsilon$ is the smallest positive distance between any two tuples in layer $l$ over some attribute (line 1). We let $T$ be the set of all such tuples (line~9); note that $|T|=3^k$, where $k$ is the number of attributes. We now find the group $g'$ (\textsf{GetGroup}$(l,t)$, line 11), which contains the constructed tuple $t$, and add its representative tuple to $S'_l$ and all its constituent tuples to $S_{l-1}$ (line~13,14). The efficiency of \neighbor critically relies on the efficiency of \textsf{GetGroup}$(l,t)$. A naive implementation of \textsf{GetGroup}$(l,t)$ would be to linearly scan all the groups to find where tuple $t$ belongs. We show in Appendix~\ref{appendix:dlv_large} that \dlv can achieve sub-linear time complexity for the function \textsf{GetGroup}$(l,t)$. This sampling of neighboring tuples continues as long as $|S_{l-1}| < \alpha$ (line~4). 

\begin{expbox}
\noindent\textbf{Mini-Experiment 2.} \textit{Does replacing \neighbor with a random sampling of representative tuples impact the overall performance of \lsr?}

We ran \common{query Q1 \ssds} described in Table \ref{tab:benchmark} with query-hardness levels $\hard \in \{1,3,5,7,9,11,13\}$ (Section~\ref{sec:Esetup}). For each $\hard$, we randomly sampled 5 sub-relations of size 10 million representing 5 queries for a total of 35 queries. We compared the results between two \lsr variants: one with \neighbor and one where \neighbor is replaced by a random sampling of tuples. \lsr with \neighbor solved all of the 35 package queries while the random-sampling variant solved one less. On average, the solvable queries showed a 7.67x improvement in the objective value when using \neighbor. \common{Experiments with other queries yield similar results; see \cite[Figure~\ref{fig:m2}]{pq} for details.}
\end{expbox}

\subsection{\pds}
\label{sec:pds}

A key ingredient in \lsr is the LP solver. Typical LP sizes range from hundreds of thousands to tens of millions of variables. The standard dual simplex algorithms in commercial systems such as \gb or \cp are sequential and make no assumptions on the number of variables versus the number of constraints. In this generic setting, prior works~\cite{dual_parallel1, dual_parallel2} have tried to efficiently parallelize dual simplex for up to 8 processing cores. Specifically, in \cite{dual_parallel1}, the authors observed a 2.34x speedup at 8 cores, with 65\% of the execution effectively parallelized.

We introduce a novel algorithm, \pds, that achieves superior speedup by exploiting the special structure of the ILPs 
that arise when solving package queries. In textbook ILPs (such as set cover \cite{set_cover}, unit commitment \cite{unit_commitment}, knapsack sharing \cite{ksp}, and traveling salesman \cite{tsp_ilp}), the number $m$ of constraints is a polynomial function of the number $n$ of variables. In contrast, a package query ILP has a constant number of constraints $m$ that is much smaller than $n$. By exploiting this structural difference, we greatly simplify our dual simplex implementation and are also able to efficiently parallelize most of the dual simplex sub-procedures. Roughly speaking, in dual simplex, we quickly move from one solution to another better one by selecting a good direction via \textit{pivoting} \cite{pivot}. Moving between solutions involves multiplications of an $n\times m$ matrix by an $m$-vector, which can be parallelized over $n$. Furthermore, the search for a good direction is a sequential operation but can be parallelized efficiently as we observed in our experiments assuming that we have a few constraints $m$ and a huge number of variables $n$. See Appendices~\ref{appendix:pds}, and \ref{appendix:pdsTricks} for technical details.

\begin{expbox}
\textbf{Mini-Experiment 3.} \common{\textit{How well does \pds scale with more cores?}}

We found that our \pds algorithm can scale up to at least 80 cores, attaining a 4.79x speedup, with 80\% of the execution effectively parallelized---a significant improvement over generic parallel dual simplex implementations. \common{See \cite[Figure~\ref{fig:m3}]{pq} for details.}
\end{expbox}

\subsection{\filp}
\label{sec:filp}

\begin{algorithm}[t]
\caption{\filp}\label{alg:filp}
\begin{algorithmic}[1]

\Require $Q := $ package query
\Statex $S := $ set of indices of $n$ tuples in the relation
\Ensure $q := $ initial size of the sub-ILP

\State $P \gets \getLP(Q[S])$
\State $x^* \gets $\pds$(P)$
\State $E \gets \sum_{i=1}^n x^*_i$
\State $P' \gets P$ where the upper bound of each variable is $E/q$
\State $y^* \gets $\pds$(P')$
\State $S' \gets \{i | x^*_i > 0 \vee y^*_i > 0\}$ 

\State $P^* \gets \getILP(Q[S'])$
\State $S^* \gets$ \textsf{
ILPSolver}$(P^*)$ 

\While{$S^*=\emptyset$ and $q < n$} \Comment{ {\color{gray}\textit{Fallback mechanism}}}
    \State $q \gets \min(2q, n)$
    \State Uniformly sample $S_u \subseteq \{i | i \notin S'\}$ such that $|S_u|=q-|S'|$
    \State $S' \gets S' \cup S_u$
    \State $P^* \gets \getILP(Q[S'])$
    \State $S^* \gets$ \textsf{ILPSolver}$(P^*)$
\EndWhile
\State\Return $S^*$

\end{algorithmic}
\end{algorithm}

\filp is a novel heuristic (Algorithm \ref{alg:filp}) for efficiently and approximately solving the final ILP encountered in \lsr (line 7 of Algorithm \ref{alg:lsr}). 
It is a type of Relaxation Enforced Neighborhood Search (RENS) heuristic \cite{rens, milp_heuristic}. RENS are characterized by constructing a sub-ILP (to be solved by a black-box ILP solver) where most of the zero decision variables in the LP relaxation $x^*$ are hard-fixed to 0. 

\smallskip\noindent\textbf{Number of positive coefficients in the LP solution.} 
\filp initially computes the LP solution $x^*$ (line~1-2). For $x^*$, the theory of the simplex method~\cite{basic_simplex} asserts that the number of \textit{basic} variables that can take fractional values is at most the number of constraints $m$. Assuming that the upper bound of each variable is 1, the number of \textit{non-basic} variables, which can either be 0 or 1, is therefore $n-m$, where $n$ is the number of variables. Letting $E=\sum_{i=1}^n x^*_i$ (line 3) be the sum of all decision variables of $x^*$, i.e. the L1 norm of $x^*$, we see that the number of variables that are 0 is at least $\lfloor n-m-E \rfloor$. Note that most of the decision variables are 0 in $x^*$ since $n \gg m+E$ and only a few variables are positive, i.e., at most $\lceil m+E \rceil$ of them. We can now use $x^*$ to construct a reduced-size sub-ILP from the positive variables. Let \defbox{$q$} be the size of this sub-ILP. 

\smallskip\noindent\textbf{Configuring $\boldsymbol{q}$.}
\rc{If $q \approx E$, i.e. we have pruned out all zero-valued decision variables, we may end up with false infeasibility (the sub-ILP is infeasible but the ILP itself is feasible) or sub-optimality.} If $q$ is too large, then we may incur unnecessary and significant computational costs. The right value of $q$ should be small enough to allow the sub-ILP to be solved within interactive performance by an off-the-shelf black-box ILP solver, (i.e. sub-second time), yet large enough to comfortably contain the typical solution sizes for package queries. E.g., package queries in our benchmark (Section \ref{sec:Esetup})  typically have solutions with 10 to 1000 tuples ($E \approx [10-1000]$), and setting \common{$q=500$} achieves the right balance of interactive performance and feasibility. \rb{See \cite[Mini-Experiment~7]{pq} for the impact of $q$ on the performance of \filp.}

\smallskip\noindent\textbf{Sub-ILP.} From $x^*$ and $q$, \filp constructs an auxiliary LP $P'$ such that its solution has approximately $q$ positive variables (lines~4-5). We observe that the sum of all decision variables of $x^*$ is $E$ when the upper bound of each variable is 1. Hence, by limiting the upper bound of each variable to $E/q$, we hope to have at least $q$ positive variables. This simple modification effectively forces the LP solver to distribute its choices evenly across the tuples to produce $q$ positive decision variables in $y^*$. \filp now formulates and solves the sub-ILP using tuples with positive coefficients in both the initial and the auxiliary LP solutions, $x^*$ and $y^*$ (lines 6-8).

\smallskip\noindent\textbf{Fallback mechanism.}
Unlike other RENS heuristics, \filp has a graceful fallback mechanism to handle false infeasibility if $q$ is insufficiently large (line 9). \filp doubles $q$ and randomly samples more tuples to include in the sub-ILP from the original relation (lines 11-12) until it includes the full relation. In practice, we observed that many of the difficult queries could be solved after one or two fallback iterations, i.e., doubling or quadrupling the initial sub-ILP size, without falling all the way back to the original relation. 

\begin{expbox}
\noindent\textbf{Mini-Experiment 4.} \textit{Does replacing the Auxilary LP with a random sampling of tuples from $S$ to formulate a sub-ILP of size $q$ impact the overall performance of \filp ?} 

We ran \common{query Q1 \ssds} described in Table \ref{tab:benchmark} with query-hardness levels $\hard \in \{1,3,5,7,9,11,13\}$ (Section~\ref{sec:Esetup}). For each $\hard$, we randomly sampled 5 sub-relations of size 1 million representing 5 queries for a total of 35 queries. We compared the results between two \filp variants: one with the Auxiliary LP $P'$ and one with a random sampling, i.e., replacing line 6 of Algorithm \ref{alg:filp} with $S' \gets \{i | x_i^*>0 \vee u_i<q/n\}$ where $u_i \sim \mathcal{U}(0,1)$. \filp with Auxiliary LP solved all 35 queries, while \filp with random sampling solved only 25. For queries solved by both variants, we observed an improvement in the objective value by 1.135x on average when using \filp with Auxiliary LP. \common{Results for other queries were similar; see \cite[Figure~\ref{fig:m4}]{pq} for details.}
\end{expbox}
\section{Partitioning algorithm}
\label{sec:dlv}

\dlv (\adlv) is a novel partitioning algorithm that works with multidimensional tuples (Section~\ref{sec:dlv_multi}) and very large relations (Appendix~\ref{appendix:dlv_large}). The algorithm relies on the \onedlv (\aonedlv) subroutine that iteratively selects and partitions a relation one attribute at a time. \aonedlv is unlike traditional partitioning algorithms such as \kd: in one iteration, it partitions an attribute using $p\ge 2$ flexible intervals instead of just two intervals separated by the attribute's mean. 

\begin{definition}[p-Partition]
\label{def:p}
Given a set $S$ of tuples, and a vector $d=(d_0,d_1,\ldots,d_p)$ where $p \geq 1$ and $-\infty = d_0 < d_1 < \ldots < d_{p-1} < d_p = \infty$, the \emph{$p$-partition} $\xP_d(S, j)$ of the set $S$ over an attribute $j$ is the disjoint partition $\{P_1,P_2,\ldots, P_p\}$ of $S$ such that $P_i=\{t \in S:d_{i-1}\le t.j < d_{i}\}$ for $1\le i\le p$, where $t.j$ is the attribute $j$ of tuple $t$.
\end{definition}

\subsection{\onedlv}
\label{sec:dlv_one}

The core idea of \aonedlv is to minimize the variance of each subset $P_i$ in a $p$-partition by dynamically allocating more $P_i$'s to partition a spread-out set of attribute values and fewer $P_i$'s to partition a concentrated one. The procedure is given as Algorithm \ref{alg:1d}.

\begin{algorithm}[t]
\caption{\onedlv}\label{alg:1d}
\begin{algorithmic}[1]

\Require $\beta := $ bounding variance
\Statex $S := $ set of $k$-dimensional tuples of size $n$ 
\Statex $j := $ attribute to partition

\State $\Tilde{S} \gets $ list of tuples in $S$ sorted in increasing order of attribute $j$

\State $V \gets \emptyset$
\State $d \gets \{-\infty, \infty\}$
\For{each $t \in \Tilde{S}$}
    \If{$\sigma^2(V \cup \{t.j\}) > \beta$} \Comment{ {\color{gray}\textit{$\sigma^2$ is the variance function}}}
        \State $d \gets d \cup \{t.j\}$
        \State $V \gets \emptyset$
    \EndIf
    \State $V \gets V \cup \{t.j\}$
\EndFor
\State $\Tilde{d} \gets $ vector of values in $d$ sorted in increasing order

\State \Return $\xP_{\Tilde{d}}(S,j)$

\end{algorithmic}
\end{algorithm}

Given a specified value $\beta > 0$ called the \textit{bounding variance}, \aonedlv iterates through the attribute values in increasing order (line~1). The algorithm keeps track of a running variance of the values grouped so far (line~9). Once this variance exceeds $\beta$ (line~5), it places a delimiter between the current tuple and the previous one and resets the running variance (lines~6-7).

\smallskip\noindent\textbf{Configuring $\boldsymbol{d_f}$.}
Recall that a smaller \df \dfv in Section \ref{sec:lsr} yields a smaller expected number of tuples in by each group. A representative tuple more accurately represents its group's tuples if the group has fewer, more concentrated tuples. We also augment the solution package $S'_l$ at every \shading iteration with neighboring representative tuples (Section \ref{sec:neighborSearch}). If we have smaller, and hence more, neighboring groups, we can add more representative tuples to $S'_l$ during Neighbor Sampling up to the \aug \augv and thus better capture hidden outliers. However, the smaller the \dfv, the higher the computational cost of \lsr as the depth of the hierarchy of relations increases. We observed that $\dfv \approx [10-1000]$ achieves the right balance between accuracy and computation cost. 

\begin{figure}
\centering
\includegraphics[width=0.85\linewidth]{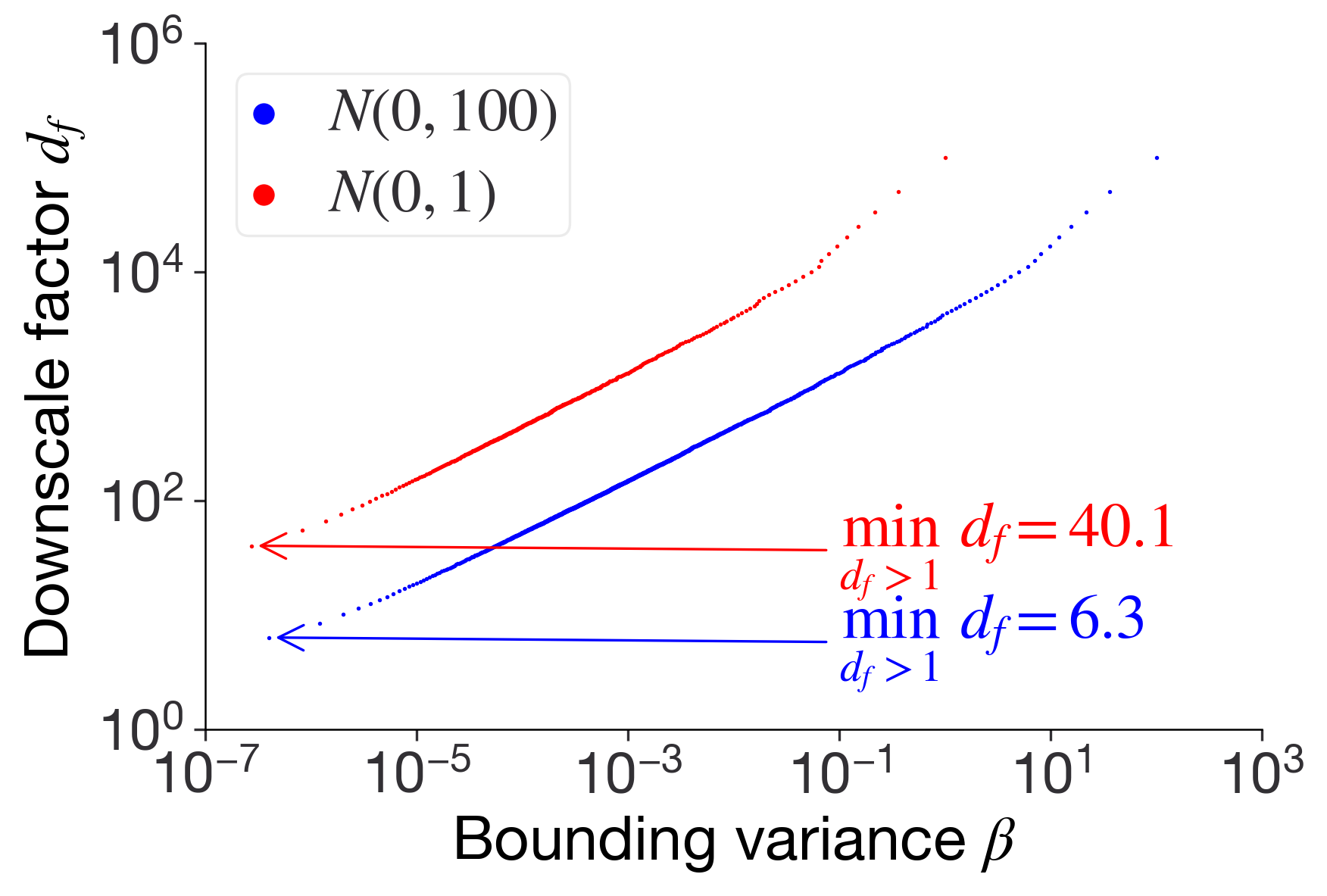}
\caption{The observed \protect\df \protect\dfv for different bounding variances $\beta$ under two normal distributions $\mathcal{N}(0,1)$ and $\mathcal{N}(0,100)$.}
\label{fig:p}
\vspace{-2mm}
\end{figure}

\smallskip\noindent\textbf{Configuring $\boldsymbol{\beta}$.} 
In \lsr, given a \df \dfv, one wishes to find a bounding variance $\beta$ such that the $p$-partition produced by \aonedlv has $p \approx n/\dfv$ where $n$ is the relation size. However, \aonedlv with a single bounding variance $\beta$ can fail to achieve certain small target values for \dfv, especially when the variance of the distribution is low; see Figure \ref{fig:p}. This is an issue since \lsr requires \dfv to be very small ($\dfv \approx [10-1000]$). \adlv overcomes this issue by using multiple bounding variances on multiple attributes and as a result extends to multidimensional settings.

\subsection{\dlv}
\label{sec:dlv_multi}

\begin{algorithm}[t]
\caption{\dlv}\label{alg:dlv}
\begin{algorithmic}[1]

\Require $S := $ set of $k$-dimensional tuples
\Statex $\dfv := $ \df
\State $(c_1,c_2,\ldots,c_k)\gets $ \textsf{GetScaleFactors}$(S, \dfv)$.

\State $\xP \gets \{S\}$ \Comment{ {\color{gray}\textit{$\xP$ is a max-priority queue}}}

\While{$|\xP| < |S|/\dfv$}
    \State $P^* \gets \argmax_{P \in \xP} |P| \max_{j}\sigma^2(P,j)$
    \State $j^* \gets \argmax_{j}\sigma^2(P^*,j)$
    \State $\beta \gets c_{j^*}\sigma^2(P^*,j^*) / \dfv^2$
    \State $\xP \gets (\xP \setminus \{P^*\}) \cup $ \textsc{\onedlv}$(\beta, P^*, j^*)$
\EndWhile

\Return $\xP$

\end{algorithmic}
\end{algorithm}

\adlv is displayed as Algorithm~\ref{alg:dlv}. It is a divisive hierarchical clustering algorithm~\cite{divisive} where all tuples start in one cluster (line~2) and splits are performed recursively until we reach $\approx |S|/\dfv$ clusters (line~3) where $|S|$ is the number of tuples. The splitting always prioritizes the cluster $P^*$ with the maximum highest total variance using a max priority queue (line~4) where $\sigma^2(P,j)$ is the variance of the attribute $j$ of tuples in $P$. For cluster $P^*$, we partition on the attribute $j$ having the highest variance (line~5). As discussed below, the bounding variance $\beta$ is set in such a way that the $p$-partition for the cluster $P^*$ produced by \aonedlv has approximately $\dfv$ subsets (lines~6-7).

Intuitively, \adlv is analogous to iteratively partitioning the terrain squares in order of highest squares first in our stylized example (Figure \ref{fig:res}) where each iteration corresponds to partitioning a square into $\dfv=4$ smaller squares. Hence, the first heuristic is to come up with a bounding variance $\beta$ (line~6) so that in each iteration, \aonedlv partitions $P^*$ into approximately $\dfv$ subsets (line~7). Let $\sigma^2$ be the variance of the partitioning attribute of $P^*$. We observed that the appropriate form for $\beta$ is $c\sigma^2 / \dfv^2$ for a constant $c>0$ since $P^*$ is partitioned into approximately $\dfv$ subsets so the variance of each subset is expected to decrease by a factor of $\dfv^2$. Moreover, the value of $c$ depends on the distribution of $P^*$ and we can accurately find such $c$, by simply binary searching $\beta$ for each $P^*$ assuming that the number of partitioning subsets of $P^*$ is a decreasing function of $\beta$. However, this approach requires us to do multiple executions of \aonedlv over $P^*$ in each iteration and hence is slow in practice. To avoid this, we can approximate $c_j$ for each attribute $j$ before the iterations via the \textsf{GetScaleFactors} function (line~1) which essentially samples the attribute values and then does a binary search. See Appendix~\ref{appendix:dlv_scale} for the details of the function. For our datasets, we found $c=13.5$ to work well.

The second heuristic is to choose a ranking that best captures the variability of a multi-dimensional subset $P \in \xP$ (line~4). For each subset of tuples, one can compute the variance or the total variance (i.e., variance times set size) for each attribute and take the maximum over all the attributes. We observed empirically that using the total variance would produce much better solutions compared to using the variance. There are several advantages to using \adlv:

\begin{itemize}[leftmargin=*]
    \item It partitions on multiple attributes and produces partitions for any given number of partitioning subsets.
    \item The actual \adlv partitioning operation is usually executed on a partition much smaller than $S$. This allows sorting algorithms on these smaller subsets to be much faster and cache-friendly.
    \item The average number of passes through the relation is $\mathcal{O}(\log_{\dfv} n/\dfv)$ where $n$ is the relation size and $\dfv$ is the \df.
\end{itemize}

\common{In Appendix~\ref{appendix:dlv_large}, we show how to extend \adlv to run on large relations via a bucketing scheme.} 

\subsection{Comparison to \kd}
\label{sec:dlv_score}

\noindent\textbf{Partitioning score.} Partitioning of a relation groups similar tuples together. A representative tuple can then be computed as an average over the similar tuples in the group. This similarity can be quantified by the tuples' distances to the representative.
If we take the average of the squared distances between these tuples and the representative then this measure corresponds to the variance of the tuples' attributes. Therefore, the variance of the tuples' attributes reflects, on average, how spread out or clustered the group's tuple attribute values are and thus the similarity of tuples within the group. A good partitioning algorithm will create groups with more tightly clustered tuples, i.e., low within-group variance. Consequently, a useful measure of how well a partitioning algorithm performs will reflect the changes in within-group variance before and after partitioning. We define the \textit{Ratio Score} as such a measure. For simplicity, we will restrict our analysis to partitions over one-dimensional tuples:

\begin{definition}[Ratio score]
For the $p$-partition $\xP_d(S)$ of the set $S$ of one-dimensional tuples, let $\sigma_i^2$ be the variance of the tuple values in partition $P_i$ ($1\le i\le p$) and $\sigma^2>0$ be the variance of tuple values in the unpartitioned set $S$. The ratio score $z(\xP_d(S))$ is $\sum_{i=1}^p \sigma_i^2 / \sigma^2$.
\end{definition}

Intuitively, the ratio score is the ratio between the sum of the subsets' variance and the set's variance. Hence, the lower the score, the better the partitioning algorithm. The lowest score is 0 when all the subsets have a variance of 0. On the other hand, if all of the subsets are empty except one $P_{i'}$ then $\sigma_i^2=0$ for $i \neq i'$ while $\sigma^2_{i'}=\sigma^2$. Hence, the score is 1 for such a trivial partition. Ratio scores that exceed 1 are possible, as shown in Theorem~\ref{the:unsorted_kd} below.

\begin{figure}
    \centering
    \includegraphics[width=0.8\linewidth]{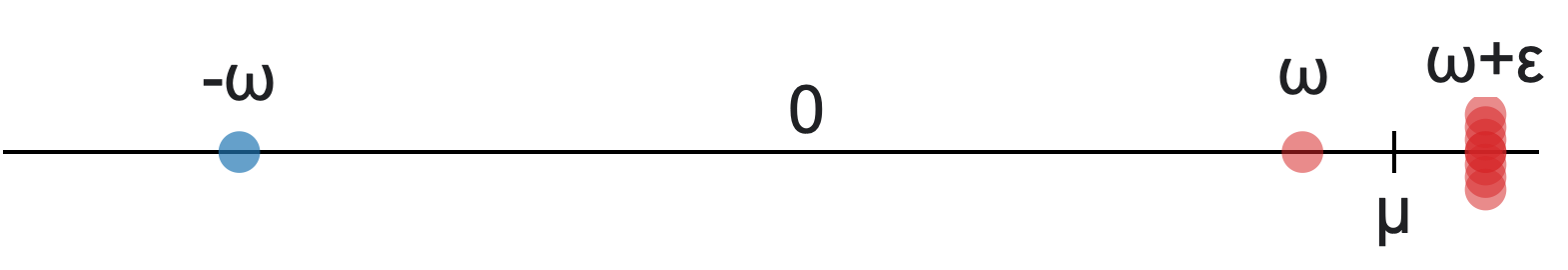}
    \vspace{-3mm}
    \caption{For a distribution consisting of one value each for $-\omega$ and $\omega$ and many values at $\omega+\epsilon$, the first \kd split is at $\mu$ between $\omega$ and $\omega+\epsilon$, forcing a grouping of the highly discrepant values $-\omega$ and $\omega$.}
    \label{fig:heavy_like}
\end{figure}

\smallskip\noindent\textbf{\kd versus \aonedlv.} \kd is also a divisive hierarchical clustering algorithm \cite{divisive} where a cluster is always split into two smaller clusters using the mean value. \ra{For generating up to several thousands of clusters, \kd is more efficient than other traditional clustering algorithms since each pass through the relation essentially doubles the number of clusters produced.
Once the number of generated clusters goes beyond millions, however, its performance deteriorates.}
In \citet{brucato}, a cluster $P_i$ is considered for splitting if it satisfies one of the two conditions: (1) its size $|P_i|$ is more than \textit{size threshold} $\tau \geq 1$; and (2) its radius $r_i$ is more than \textit{radius limit} $\omega \geq 0$ where $r_i=\max_{x \in P_i} |x-\mu(P_i)|$ and $\mu(P_i)$ is the mean of $P_i$. The following result shows that on some data sets the clustering performance of \kd degrades totally while \aonedlv attains almost perfect clustering.

\begin{theoremrep} For any radius limit $\omega>0$, there exists a sequence $\{S_n\}$ of sets of one-dimensional tuples whose variances converge to 0 such that for any size threshold $\tau \geq 2$, \kd's ratio score tends to $\infty$ as $n \to \infty$. On the other hand, using \aonedlv with a bounding variance $\beta=24\sigma^2(S_n)/|S_n|^2$, the ratio score converges to 0.
\label{the:unsorted_kd}
\end{theoremrep}

\begin{appendixproof}

Let $S_n$ be a set of $n+2$ tuples which has 1 value of $-\omega$, 1 value of $\omega$ and $n$ values of $\omega+\epsilon$ where $\epsilon=3\omega/n$.

Denote $\mu(S_n)$ and $\sigma^2(S_n)$ be the mean and the variance of $S_n$ respectively. Then

\[
\mu(S_n)=\frac{n(\omega+\epsilon)}{|S_n|}=\frac{n+3}{n+2}\omega>\omega
\]

Hence $\mu(S_n)$ is between $\omega$ and $\omega+\epsilon$. Thus, \kd first splits $S_n$ into $P_1$ and $P_2$ using $\mu(S_n)$ where $P_1$ contains two tuples of values $-\omega$ and $\omega$ and $P_2=S_n \setminus P_1$. The radius of $P_1$ is $\omega$ and hence $P_1$ is not considered for splitting. Moreover, since $P_2$ contains tuples whose values are $\omega+\epsilon$, all the subsequent splits from $P_2$ have subsets of variance 0. Therefore, for the $p$-partition produced by \kd, $\sum_{i=1}^p \sigma^2(P_i)=\sigma^2(P_1)=\omega^2$.

To compute the ratio score, we first compute the variance of $S_n$:

\[
\begin{split}
&\sigma^2(S_n)
=\mathbb{E}[S_n^2]-\mu(S_n)^2\\
&\ =\frac{2\omega^2+n(\omega+\epsilon)^2}{|S_n|}-\frac{n^2(\omega+\epsilon)^2}{|S_n|^2}\\
&\ =2\omega^2\left[\frac{1}{n+2}+\frac{(n+3)^2}{n(n+2)^2}\right]
\end{split}
\]

Observe that $\lim_{n\to \infty} \sigma^2(S_n) = 0$. Hence, $z(\xP_d(S_n))=\frac{\sum_{i=1}^p \sigma^2(P_i)}{\sigma^2(S_n)}=\frac{\omega^2}{\sigma^2(S_n)}$ goes to infinity as $n \to \infty$. This proves the first assertion. 

To prove the second assertion, consider the $p$-partition $\xP_d(S_n)$ using \aonedlv with a bounding variance
\[
\beta=24\frac{\sigma^2(S_n)}{|S_n|^2}=48\omega^2\left[\frac{1}{(n+2)^3}+\frac{(n+3)^2}{n(n+2)^4}\right]
\]
Since $\sigma^2(\{-\omega,\omega\})=\omega^2>\beta$
for $n \geq 4$, we have that $P_1$ contains only one tuple of value $-\omega$. In addition, since
\[
\sigma^2\bigl(\{\omega,\omega+\epsilon\}\bigr)=\frac{1}{4}\epsilon^2=\frac{9}{4}\frac{\omega^2}{n^2} > \beta
\]
for $n \geq 39$, it follows that $P_2$ contains only one tuple of value $\omega$. Hence, for \aonedlv, the first two values $-\omega$ and $\omega$ are isolated from each other so that all the variances of $P_i$ are zero. Therefore, for $n \geq 39$, the ratio score for \aonedlv is 0 and the result follows.
\end{appendixproof}

We include the full proofs of our theoretical results in Appendix~A. To prove Theorem \ref{the:unsorted_kd}, we construct a sequence $\{S_n\}$ of sets of one-dimensional tuples as in Figure \ref{fig:heavy_like}. We force \kd to group two very dissimilar values by exploiting the fact that the splitting intervals of \kd are fixed as long as the mean of the values does not change. \aonedlv, on the other hand, with an appropriate choice of bounding variance, overcomes this issue. Indeed, we now show that \aonedlv has a low ratio score for virtually any large relation. Specifically, for any set of $n$ one-dimensional tuples, \aonedlv, using the above bounding variance, achieves an $O(1/n)$ ratio score. Moreover, the corresponding partitioning is nontrivial in that there exist partitions with at least two tuples, i.e., $p<n$.

\begin{theoremrep}[Universal bounded ratio score]
Let $S$ be a set of one-dimensional tuples of size $n\geq 2$ with variance $\sigma^2 > 0$. Then \aonedlv with a bounding variance $\beta=24\sigma^2/n^2$ will produce a $p$-partition $\xP_d(S)$ where $p \leq (3/4)n+1/2$---so that the partitioning is nontrivial---and $z(\xP_d(S)) \leq 24/n$.
\label{the:bounded}
\end{theoremrep}
\begin{appendixproof}

Let $S$ be a set of $n$ one-dimensional tuples whose values are $x_1 \leq x_2 \leq ... \leq x_n$. Consider $n-1$ intervals $[x_i, x_{i+1}]$ where $i=1,2,...,n-1$. The length of each interval is simply $\lvert x_i-x_{i+1}\rvert=x_{i+1}-x_i$. Let $s \in [0,n-1]$ be the number of intervals whose length is greater than $2\sqrt{\beta}=4\sqrt{6}\sigma/n$. We call these \textit{critical intervals}. 

We first claim that $s$ is at most $n/2$. To prove our claim, we assume without loss of generality that $s \geq 2$; if $s<2$ then we are done. Denote by $\mu$ the mean of the values in $S$ and observe that each of the critical intervals either lies on the left of $\mu$ or the right of $\mu$ or contains $\mu$. Let $l,r$ be the number of critical intervals that lies completely on the left and the right of $\mu$ respectively. Observe that there are at least $l$ endpoints such that distance from the $i$th endpoint to  $\mu$ is $\ge 2i\sqrt{\beta}$ for $i\in[1..l]$. Let $g(x)=\bigl(2x(x+1)(2x+1)\beta\bigr)/3$ and $A_L$ be the sum of squares of those distances so that $A_L \geq \sum_{i=1}^l (2i\sqrt{\beta})^2=4\beta \sum_{i=1}^l i^2=g(l)$. We analogously define $A_R$ where $A_R \geq g(r)$. Because the variance $\sigma^2$ of $S$ is the mean of the squared distances between the $x_i$'s and $\mu$, we have
\[
\sigma^2 \geq \frac{1}{n}(A_R+A_L) \geq \frac{1}{n}\bigl(g(r)+g(l)\bigr) \geq \frac{2}{n} g\bigl(\frac{r+l}{2}\bigr)
\]
by Jensen's inequality since $g(x)$ is a convex function when $x \geq 0$. For the case $l+r=s$, i.e., $\mu$ does not lie in any of these critical intervals, then
\[
\sigma^2 \geq \frac{2}{n}g\bigl(\frac{s}{2}\bigr)=\frac{\beta}{3n}s(s+1)(s+2) > \frac{\beta}{3n}s^3.
\]
For the case where $l+r=s-1$, i.e., where $\mu$ lies in one of these critical intervals, let $a_L,a_R$ be the distance from $\mu$ to the left and right endpoints of the critical interval that contains $\mu$ and observe that $a_L+a_R > 2\sqrt{\beta}$. Also set $h(x,y)=xy^2$ and note that $h$ is a convex function when $x,y \geq 0$ and an increasing function in $x$ and $y$ respectively. Observe that the $l$ critical intervals that lie completely on the left of $\mu$ are now essentially shifted to the left by a distance $a_L$ compared to the first case. Hence
\[
\begin{split}
A_L &\geq \sum_{i=1}^l (2i\sqrt{\beta}+a_L)^2
\geq \sum_{i=1}^l (2i\sqrt{\beta})^2 + a_L^2
=g(l)+la_L^2\\
&=g(l)+h(l,a_L).
\end{split}
\]
Similar reasoning shows that $A_R \geq g(r)+h(r,a_R)$. Putting these results together, we find that
\[
\begin{split}
\sigma^2 &\geq \frac{1}{n}(A_L+A_R)\\
&\geq \frac{1}{n}\bigl(g(l)+g(r)+h(l,a_L)+h(r,a_R)\bigr)\\
&\geq \frac{1}{n}\Bigl(2g\bigl(\frac{l+r}{2}\bigr)+2h\bigl(\frac{l+r}{2}, \frac{a_L+a_R}{2}\bigr)\Bigr)\\
&> \frac{1}{n}\Bigl(2g\bigl(\frac{s-1}{2}\bigr)+2h\bigl(\frac{s-1}{2},\sqrt{\beta}\bigr)\Bigr)\\
&=\frac{1}{n}\bigl(\frac{\beta}{3}(s^3-s)+(s-1)\beta\bigr)\\
&=\frac{1}{n}\bigl(\frac{\beta}{3}s^3+\frac{2\beta}{3}s-\beta\bigr)\\
&>\frac{\beta}{3n}s^3
\end{split}
\]
since $s \geq 2$. Hence, in both cases, we have proved that
\[
\sigma^2 > \frac{\beta}{3n}s^3=8\frac{\sigma^2}{n^3}s^3,
\]
which implies that $s^3 < n^3/8$ and hence $s < n/2$ as claimed.

Now consider the execution of DLV with bounding variable $\beta$. Denote by $E_i$ the event in which the current running variance exceeds $\beta$ and forces DLV to delimit between the current element $x_i$ and the previous element $x_{i-1}$. Event $E_1$ happens as we start. We claim that after the partitioning is done, if $P_i=\{x_j\}$ for some $j<n$ then $[x_j,x_{j+1}]$ is a critical interval. To see this, observe that since $x_j$ is not the last element $(j<n)$ then both $E_{j}$ and $E_{j+1}$ have happened. Once $E_j$ happens, we know that $P_i=\{x_j\}$. However, since $E_{j+1}$ has also happened, the variance of $\{x_j,x_{j+1}\}$ must have been greater than $\beta$ which implies that the distance between $x_{j+1}$ and $x_j$ must have been greater than $2\sqrt{\beta}$, hence implying that $[x_j,x_{j+1}]$ is a critical interval. 
Assume that we have more than $s+1$ subsets $P_i$ such that $|P_i|=1$, then from $x_1,...,x_{n-1}$ (excluding that last element), we would have more than $s$ subsets $P_i$ such that $|P_i|=1$. This implies that we would have more than $2s$ endpoints (repetitions counted) that belong to some critical intervals. However, given only $s$ critical intervals, one cannot have more than $2s$ such endpoints. Therefore, we can have at most $s+1$ subsets $P_i$ such that $|P_i|=1$. Then at least $n-s-1$ points belong to some subset $P_i$ where $|P_i| \geq 2$. Hence the number $p$ of subsets $P_i$ satisfies
\[
p \leq s + 1 + \big\lfloor\frac{n-s-1}{2}\big\rfloor \leq \frac{n+s+1}{2} < \frac{n+n/2+1}{2}=\frac{3}{4}n+\frac{1}{2}.
\]
By construction, \onedlv ensures that $\sigma^2_i\le\beta$ for all $i$ and thus we know that
\[
z\bigl(\xP_d(S)\bigr) \leq \frac{p\beta}{\sigma^2} \leq \frac{(\frac{3}{4}n+\frac{1}{2}) 24 \frac{\sigma^2}{n^2}}{\sigma^2} \leq 24\left(\frac{3}{4n}+\frac{1}{4n}\right)=\frac{24}{n}
\]
and the desired result follows.
\end{appendixproof}

At a high level, our proof proceeds by estimating the number of so-called \textit{critical intervals}, which are intervals of consecutive values in increasing order such that the two endpoints of any such intervals cannot be in one partitioning subset as it would violate the above bounding variance $\beta$. This, in turn, allows us to upper-bound the number of partitioning subsets containing a single value, i.e., not all partitioning subsets will contain a single value and thus upper-bound $p$ as well. As a result, the ratio score is bounded by $24/n$ where $n$ is the number of tuples.

\begin{figure}
    \centering
    \includegraphics[width=0.4\textwidth]{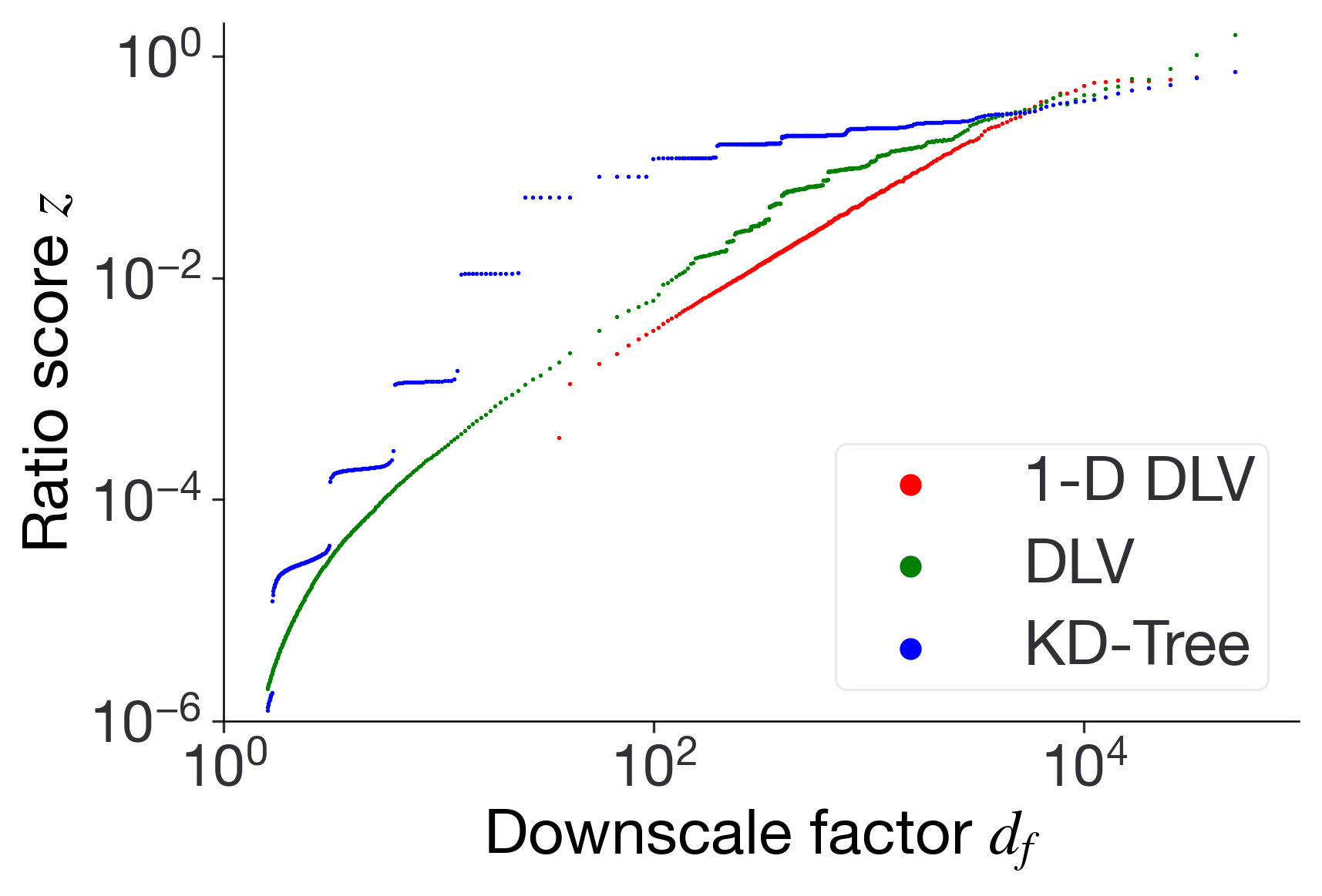}
    \vspace{-5mm}
    \caption{\adlv outperforms \kd and performs as well as \aonedlv for various values of \protect\df.}
    \label{fig:vrac}
    \vspace{-2mm}
\end{figure}

\smallskip\noindent\textbf{\adlv in practice.} Figure \ref{fig:vrac} shows the ratio score $z$ of various algorithms using the same \df $\dfv$ partitioning on a normal distribution $\mathcal{N}(0,1)$ with $10^5$ samples.

\begin{expbox}
\noindent\textbf{Mini-Experiment 5.} \textit{How efficient is \adlv compared to \kd when producing a large number of groups?} 

We ran \adlv and the \kd implementation as in \cite{brucato} to partition the dataset \tpch described in Section \ref{sec:Esetup}. \adlv partitioned a relation of $10^8$ tuples in 138s using 80 cores to produce approximately $10^6$ groups while \kd executed in 300s to produce approximately $10^3$ groups. (\kd is not well-suited to produce as many groups as \adlv due to efficiency issues as well as the inability to directly control the number of groups produced). For a relation of $10^9$ tuples, \adlv took 1827s to produce approximately $10^7$ groups while \kd ran out of memory.
\end{expbox}
\section{Evaluation}

In this section, we demonstrate experimentally that \lsr is very effective at overcoming the false infeasibility issues of the prior art (i.e., failing to derive a solution for feasible queries) while achieving superior scalability.  We first describe the setup of our evaluation, including datasets, queries, and metrics, and then proceed to showcase our results.

\subsection{Experimental Setup}
\label{sec:Esetup}

\smallskip\noindent\textbf{Software and platform.} We use \pg v14.7 for our experiments to use built-in features such as range types to store \adlv's partitioning information and GiST indexes over these range types. The main algorithms are implemented in \textsc{C++17}, which uses the \textit{libpq} library as an efficient API to communicate with \pg and the \textit{eigen} library for efficient vector/matrix operations. For parallel implementation, we use \textsc{C++} OpenMP for multi-processing computation \cite{openmp}. For solving a sub-ILP in \filp, we use \gb v9.5.2 as our black-box ILP solver \cite{gurobi}. We run all experiments on a server with Intel(R) Xeon(R) Gold 6230 CPU @ 2.10GHz with 377GB of RAM and 80 physical cores, running on Ubuntu 20.04.4 LTS. Our implementation is available at \cite{pq}.

\smallskip\noindent\textbf{Datasets.} We demonstrate the performance of our algorithms using both real-world and benchmark data. The real-world dataset consists of 180 million tuples extracted from infrared spectra (APOGEE/APOGEE-2) of the Sloan Digital Sky Survey (\ssds) \cite{ssds}. For the benchmark dataset, we use the LINEITEM table from \tpch V3 \cite{tpch} with a scale factor of 300; the table contains 1.8 billion tuples. In order to make results comparable across the two datasets, we use the same query structure with constraint bounds that are set to achieve a specific query hardness level given the mean and standard deviation of the attributes in the dataset.

\smallskip\noindent\textbf{Queries.} Given a dataset, we have developed a novel method for systematically generating queries of varying hardness, rather than generating queries in an ad hoc manner. This approach allows comprehensive benchmarking and yields a better empirical assessment of the generalizability of a technique to other data sets or package queries, which can be arbitrarily easy or hard. Specifically, we use a package query template and systematically vary the constraint bounds to expand or shrink its feasibility region. As a simple example, a package query with the constraint $\sum_j x_j < b$ is 
trivially feasible for $b = \infty$ and infeasible for $b < 1$ (since the $x_i$'s are integer). The complexity of finding a solution also depends on the objective function and the shape of the feasible region.

\smallskip\noindent\textbf{Query Hardness.}
To precisely define query hardness, let $\mathcal{E}$ be the expected \emph{package size}, i.e., the expected number of tuples in the solution to a package query.
Without loss of generality, consider a constraint, $C_i$ of the form $\sum_j a_{ij} x_j < b_i$. Suppose attribute $A_i$ is a random variable with mean $\mu$ and variance $\sigma^2$. Then with a large enough $\mathcal{E}$ and by the central limit theorem, $\mathcal{E}^{-1}\sum_{j=1}^{\mathcal{E}}A_{i}$ follows a normal distribution $\mathcal{N}(\mu,\sigma^2\mathcal{E}^{-1}).$ Constraint $C_i$ can be reformulated as $\mathcal{E}^{-1}\sum_{j=1}^{\mathcal{E}}A_i < b_i/\mathcal{E}$. The probability, $P(C_i)$, that a random sample of $\mathcal{E}$ tuples satisfy $C_i$ is simply given by the cumulative distribution function (CDF) of the normal distribution $\mathcal{N}(\mu,\sigma^2\mathcal{E}^{-1})$ evaluated at $b_i/\mathcal{E}$. With $m$ constraints, $C_1, ..., C_m$, and assuming the attributes are independent \common{for the sake of simplicity}, the probability that a random sample of $\mathcal{E}$ tuples satisfy all the constraints is $P(C_1, C_2, ... C_m) = \prod_i^m P(C_i).$ Since the chances of satisfying a harder query's constraints with a random sample of tuples are much lower, we can define \textit{hardness} as follows: $\hard:=-\log_{10}\prod_i^m P(C_i)$. Given a template package query with constraints $C_1,\ldots, C_m$, their bounds $b_1,\ldots, b_m$ as parameters, and the expected package size, $\mathcal{E}$, we can now instantiate a specific query of a specified hardness by setting the bounds accordingly. In particular, we can set $P(C_1) = P(C_2) = \cdots =P(C_m) = 10^{-\hard/m}$ and invert the CDF function to derive the bound $b_i$ for which $P(C_i) = 10^{-\hard/m}$.

Table \ref{tab:benchmark} provides information on the underlying data distribution statistics of both data sets, the package query templates, and the bounds set for query instances of a particular hardness level $\hard$, where $\hard \in\{1,3,5,7\}$.

\begin{table*}[t]
\begin{adjustbox}{max width=\textwidth, center}
\begin{tabular}[t]{@{}ccc|ccccc|ccc|ccccc@{}}
\toprule
\multicolumn{8}{c|}{\common{Q1 \ssds}}
& \multicolumn{8}{c}{\common{Q2 \tpch}}
\\ \midrule

\multicolumn{8}{l|}{
{\small
\begin{tabular}[t]{@{}l@{}}
{\tt SELECT  \textbf{PACKAGE}(*) AS P FROM sdss R \textbf{REPEAT} 0}\\ 
{\tt \textbf{SUCH THAT}   15 $\leq$ COUNT(P.*) $\leq$ 45 AND}\\  
{\tt SUM(P.j) $\geq b_1$ AND SUM(P.h) $\leq b_2$ AND}\\     
{\tt SUM(P.k) BETWEEN $b_3$ AND $b_4$} \\ 
{\tt \textbf{MINIMIZE} SUM(P.tmass\_prox)}
\end{tabular}}
} & 
\multicolumn{8}{l}{
{\small
\begin{tabular}[t]{@{}l@{}}
{\tt SELECT \textbf{PACKAGE}(*) AS P FROM tpch R \textbf{REPEAT} 0}\\ 
{\tt \textbf{SUCH THAT}   15 $\leq$ COUNT(P.*) $\leq$ 45 AND}\\   
{\tt SUM(P.quantity) $\geq b_1$ AND SUM(P.discount) $\leq b_2$ AND}\\ 
{\tt SUM(P.tax) BETWEEN $b_3$ AND $b_4$}\\ 
{\tt \textbf{MAXIMIZE} SUM(P.price)}
\end{tabular}}
} \\

\midrule

\textbf{Attribute} & $\mu$ & $\sigma$ & $\hard$: & 1 & 3 & 5 & 7 &
\textbf{Attribute} & $\mu$ & $\sigma$ & $\hard$:  & 1 & 3 & 5 & 7  \\

{\tt tmass$\_$prox} & 14.45 & 14.96 &  &  &  &  &  & 
{\tt price} & 38240 & 23290 &  &  &  &  & \\

{\tt j} & 14.82 & 1.562 & $b_1$ & 445.37 & 455.56 & 461.91 & 466.86 & 
{\tt quantity} & 25.50 & 14.43 & $b_1$ & 772.11 & 866.29 & 924.88 & 970.61  \\

{\tt h} & 14.05 & 1.657 & $b_2$ & 420.68 & 409.87 & 403.14 & 397.89 & 
{\tt discount} & 1912 & 1833 & $b_2$ & 56456.81 & 44493.54 & 37051.09 & 31242.12  \\

{\tt k} & 13.73 & 1.727 & $b_3$ & 406.04 & 410.71 & 411.64 & 411.84 & 
{\tt tax} & 1530 & 1485 & $b_3$ & 40864.32 & 44877.91 & 45680.35 & 45852.68  \\
 
 &  &  & $b_4$ & 417.76 & 413.09 & 412.16 & 411.96 &  &  &  
 & $b_4$ & 50935.68 & 46922.09 & 46119.65 & 45947.32  \\

\bottomrule
\end{tabular}%
\end{adjustbox}
\caption{\common{Experimental Benchmark Q1 \ssds and Q2 \tpch: The package query templates, underlying data statistics, and constraint bounds at different query-hardness ($\hard$) levels.}}
\label{tab:benchmark}
\end{table*}
\vspace{-0.1cm}

\smallskip\noindent\textbf{Approaches.} 
Our evaluation contrasts three approaches:
\begin{itemize}[topsep=1pt,leftmargin=*]
    \item \gb ILP solver \cite{gurobi}:  This is a state-of-the-art solver that computes solutions to the ILP problem directly, without any considerations of partitioning. It provides the gold standard with respect to accuracy but struggles to scale to large data sizes.
    \item \sr \cite{brucato}: The prior state-of-the-art in package query evaluation employs a data partitioning and divide-and-conquer strategy to achieve scalability.
    \item \lsr: Our approach employs a multi-layer partitioning strategy that smartly \emph{augments} the size of ILP subproblems to avoid false infeasibility, and a novel mechanism to parallelize and reduce solving time.
\end{itemize}

\smallskip\noindent\textbf{Metrics.} 
We evaluate the efficiency and effectiveness of all methods. The first metric is \textit{running time}. We measure the wall-clock time to generate a solution for each method. This includes the time taken to read data from \pg and the time taken for the method to produce the solution. \rb{In particular, for \lsr and \sr, the running time is computed when running 80 cores in parallel. (We use the parallel version of \sr described in \cite{brucato}.) For \gb, only sequential execution using 4 cores is available. Therefore, we also include the running time of \lsr using 4 cores in parallel.} We limit the maximum running time of any method to 30 minutes. If a method fails to produce a solution within this time limit, it is registered as a failed run, i.e., no solution found.

\looseness-1
The second metric is the \textit{integrality gap}. Recall that the solution of the LP relaxations of an ILP is readily available because we can efficiently solve the LP problem using the Simplex algorithm~\cite{Simplex}. Hence, we use the LP objective value as the upper bound for an ILP solution in a maximization problem, and as the lower bound in a minimization problem. The \emph{integrality gap} for maximization is then defined as the ratio ILP objective over LP objective: $(Obj_{ILP}+\epsilon)/(Obj_{LP}+\epsilon)$
where $\epsilon=0.1$ is required to avoid numerical instability when $\left|Obj_{LP}\right|$ is too small.
For minimization, we simply invert the ratio. Therefore, it is always the case that the integrality gap is at least 1 assuming the objective is always positive.

\smallskip\noindent\textbf{Hyperparameters.} We set hyperparameters as follows.
\begin{itemize}[topsep=1pt,leftmargin=*]
\item\textit{\gb's MIP gap}: We keep the default value of $0.1\%$. \gb will terminate when the gap of the lower and upper bound of the optimal objective value is less than $0.1\%$ of the incumbent objective value. 

\item\textit{\sr's partitioning size threshold}: 
We find that the default setting proposed by \sr~\cite{brucato} ($10\%$ of the relation size, or $\approx 10$ partitions) results in infeasibility in the sketch phase for all queries with hardness $\hard>2$ in our benchmark. We instead set the threshold to $0.1\%$. This increases the number of partitions ($\approx 1000$) allowing for smaller groups with more similar tuples and better representatives leading to a higher solve rate, without degrading the performance of the \kd index.

\item\textit{\lsr's \protect\aug \protect\augv and \protect\df \protect\dfv}: Using grid search, we find $\augv=100{,}000$ and $\dfv=100$ to be optimal. Lower $\dfv$ would cause the partitioning time to be much greater \ra{(almost 3x longer)} while higher $\dfv$ would cause the partitioning groups to be less accurate since each group would contain more tuples. Higher $\augv$ significantly increases the query time with marginal gains to solution quality. \ra{Lower $\augv$ results in a significant drop in optimality (3x worse). Case-by-case parameter tuning can be used, if needed, incurring a tuning overhead of up to 2 hours. However, the results in Section~\ref{sec:results} indicated that the foregoing hyperparameter configuration works well with various query structures and hardnesses.} \common{See \cite[Mini-Experiment~6]{pq} for details of the grid search.}
\end{itemize}

\subsection{Results}
\label{sec:results}

\begin{figure*}[t]
    \centering
    \includegraphics[width=1\textwidth]{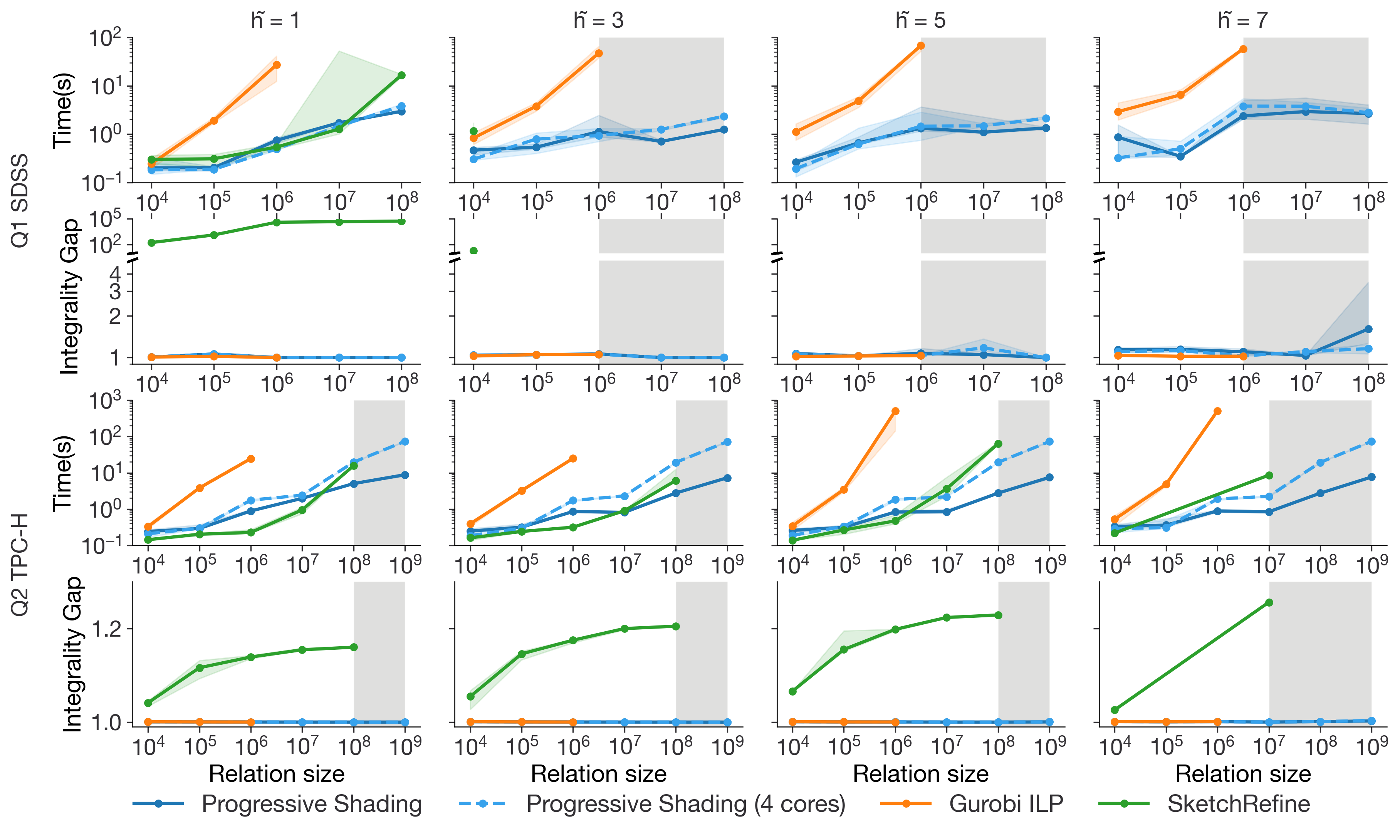}
    \vspace{-6mm}
    \caption{\common{Query performance as relation size increases for Q1 \ssds and Q2 \tpch. The point values represent the median of 10 runs and the error bands are the interquartile range (IQR) of the 10 runs.}}
    \label{fig:a4}
    \vspace{-2mm}
\end{figure*}

\noindent\textbf{Query performance as relation size increases.} \looseness-1
Figure~\ref{fig:a4} demonstrates the performance of each method as the relation size increases, as well as the effect of increasing hardness on the running time and the integrality gap. The set of hardness values that we experimented with encompasses easy to average difficulty ($\hard \in \{1,3,5,7\}$).
We generate \common{ten} relation instances for each relation size by sampling independent sub-relations from the original dataset.

In both queries, \gb only scales up to a size of one million tuples, and the running time grows exponentially with the relation size. \sr scales relatively well up to ten million tuples, but cannot sustain scaling beyond that, since the size of refined queries increases linearly with the relation size. In addition, \sr fails to find solutions for $\hard \geq 3$ for Q1 \ssds and $\hard=7$ for Q2 \tpch. In contrast, \lsr always finds solutions in both queries and achieves well below 5s running time even for one billion tuples. 

In terms of the integrality gap, \lsr achieves close-to-optimal solutions for Q1 \ssds and Q2 \tpch as seen by its integrality gap curve staying as close to that of \gb. \sr, on the other hand, produces solutions with 20\% worse objective in Q2 \tpch. For Q1 \ssds, the extremely high value of integrality gap for \sr is due to the fact that the objective column \textit{tmass\_prox} in \ssds has many zero values. This produces an LP solution with an objective value of 0. If \sr only finds an ILP solution with a positive objective value, this value will be divided by $\epsilon=0.1$, i.e., will be scaled by a factor of 10.

\smallskip\noindent\textbf{False infeasibility as hardness increases.} 
We next examine the occurrence of false infeasibility in \lsr and \sr as the hardness level becomes very high, shrinking the feasible region ($\hard$ up to 15 in our benchmark). For each query, we generate ground truth feasibility by running \gb on the query with its objective function removed. This will allow \gb to terminate as soon as it finds a feasible solution. 
We restrict the relation size to one million since this is the maximum size that \gb can solve within the time limit. Furthermore, for each dataset and hardness level, we randomly sample 20 sub-relations of size one million representing 20 queries and compute the number of queries for which each of the methods can find a solution. 

\begin{figure}[t]
\centering
\includegraphics[width=\columnwidth]{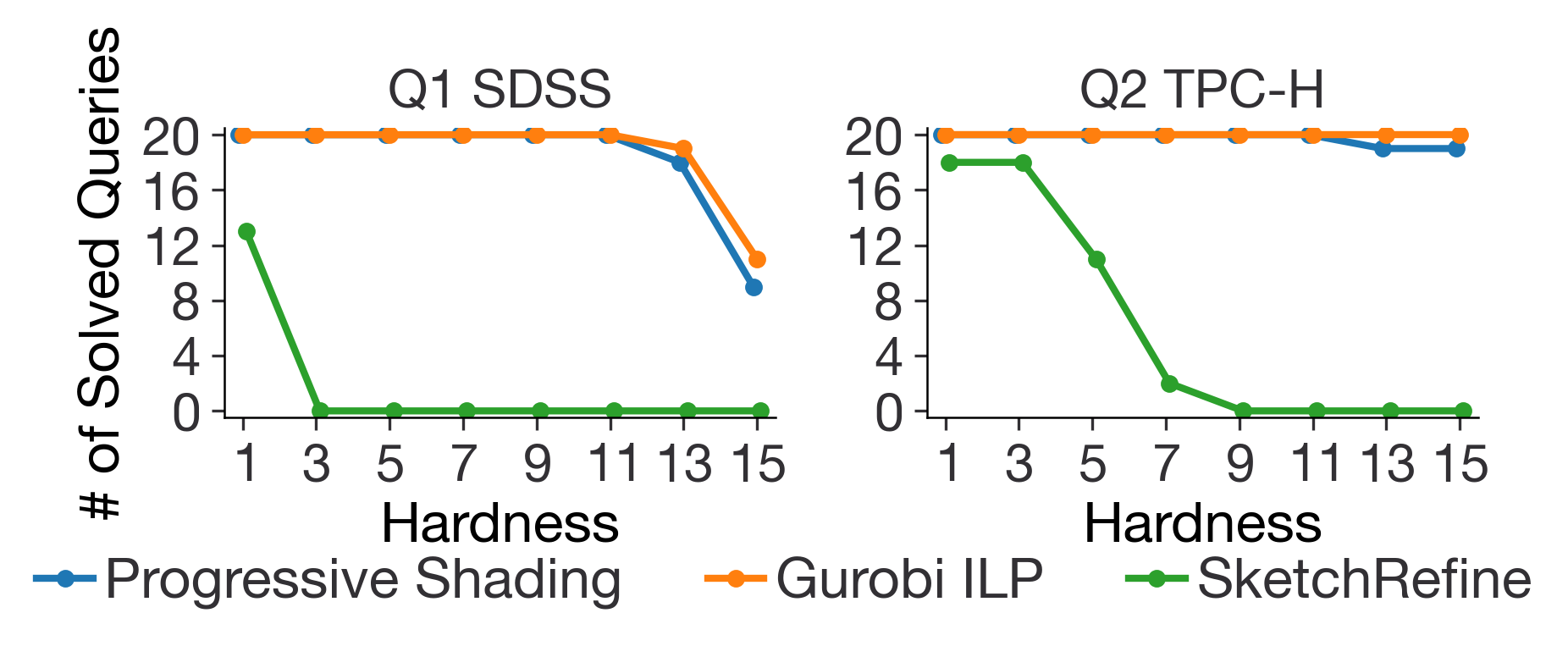}
\vspace{-6mm}
\caption{\common{False infeasibility as hardness increases.}}
\label{fig:a3}
\vspace{-4mm}
\end{figure}

\looseness-1
Figure~\ref{fig:a3} displays our results. For Q1 \ssds, \sr solves 13 out of 20 queries solved by \gb at $\hard=1$ and none at all for $\hard>1$. For Q2 \tpch, \sr solves only half as much for the usual workload of $\hard \in \{1,3,5\}$ but then fails to solve most of the hard queries where $\hard>5$. On the other hand, \lsr can solve almost as many as \gb solves in both queries. \common{Results for the other queries we examined are similar; see Appendix~\ref{appendix:ae} for details.} 

\vspace{-0.1cm} 
\section{Related work}

\noindent

\noindent\textbf{In-database optimization.} Recent research aims to integrate complex analytics capabilities into DBMSs. SolveDB \cite{solvedb} provides extensible infrastructure for integrating a variety of black-box optimization solvers into a DBMS, whereas \lsr focuses on ILP solvers and ``opens up the black box'' in order to scale to large problems. SolveDB offers built-in problem partitioning which is only applicable when there are sub-problems that can be solved independently, i.e., constraints that only exist within each sub-problem. However, such a partitioning strategy is ineffective for solving package queries because most of the tuples can connect via a single constraint and thus cannot be partitioned further. \adlv provides a simple solution to the specific needs of package queries by partitioning a very large relation into similar tuples.

\noindent\textbf{Resource allocation problems.} Partitioned Optimization Problems (POP) \cite{pop} is a recent technique to solve large-scale ``granular'' resource allocation problems that can be often formulated as ILPs whose structures are different from ILPs formulated by package queries, i.e., the number of constraints in POP can be as large as the number of variables \cite{alloc}. POP achieves high scalability by randomly splitting the problem into sub-problems and aggregating the resulting sub-allocations into a global allocation---an approach similar to \sr \cite{brucato}. Thus, POP still suffers from the same disadvantages as \sr when we increase the scale because the number of sub-problems is up to 32 in POP. Moreover, the partitioning in POP is online while \lsr is a large-scale package query solver that runs on an offline partition produced by \adlv.

\noindent\textbf{Semantic window queries.} Semantic windows \cite{window1} are related to packages. A semantic window refers to a subset of a grid-partitioned space that is contiguous and has certain global properties. For example, astronomers may divide the night sky into a grid and search for areas where the overall brightness exceeds a particular threshold. Semantic windows can be expressed by package queries with a global condition to ensure that all cells in the package are contiguous. Searchlight~\cite{window2}, a recent method for answering semantic window queries, uses in-memory synopses to quickly estimate aggregate values of contiguous regions. This approach is analogous to our hierarchical partitioning strategy using \adlv where a relation in layer~$l$ aggregates tuples from a relation in layer~$l-1$. However, Searchlight enumerates \emph{all} of its feasible solutions and retains the best one---a very expensive computation---whereas \lsr efficiently finds potentially optimal solutions via LP. 

\noindent\textbf{Neural Diving.} Neural Diving \cite{neural} is a machine learning-based approach to solving ILPs that trains a deep neural network to produce multiple partial assignments of variables in the input ILP, with the remaining unassigned variables defining smaller sub-ILPs that can be solved using a black-box ILP solver. The neural network is trained on all available feasible assignments to give a higher probability to the ones that have better objective values instead of only the optimal ones, which can be expensive to collect. The authors of \cite{neural} evaluate  the method on diverse datasets containing large-scale MIPs from real-world applications such as Google Production Planning, Electric Grid Optimization \cite{unit_commitment}, and so on. 

Unlike Dual Reducer, Neural Diving does not prune variables using an auxiliary LP but instead uses a pre-trained neural network. This approach requires expensive training over a large dataset of similar problem instances in order to learn effective heuristics. Moreover, solving package queries beyond millions is currently out of reach for Neural Diving since it requires the neural network---whose size scales with the number of variables and constraints---to fit in memory.
\section{Conclusions and Future work}

In this paper, we expand our ability  significantly beyond prior art~\cite{brucato} to solve challenging package queries over very large relations. Our novel  \lsr strategy uses a hierarchy of relations created via a sequence of partitionings, smartly \emph{augments} the size of ILP subproblems to avoid false infeasibility, and provides a novel mechanism to parallelize and reduce solving time.

In future work, we plan to investigate combining Neural Diving and \lsr to potentially solve a wide range of ILP problems (not just package queries) in arbitrarily large relations. Although Neural Diving is not currently a feasible approach, 
running large-scale neural networks inside a DBMS will eventually become efficient, e.g., by integrating tensor technology into DBMS \cite{tqp, tqp2}. Then---because it is straightforward to generate different package queries with various hardnesses and query structures---a potential approach for solving package queries with high hardness would train Neural Diving using the feasible solutions generated from \filp.

\smallskip\noindent\textbf{Acknowledgements.} This work was supported by the ASPIRE Award for Research Excellence (AARE-2020) grant
AARE20-307 and NYUAD CITIES, funded by Tamkeen under the Research Institute Award CG001, and by the National Science Foundation under grants 1943971 and 2211918.
\begin{toappendix}
\section{Linear programming in our setting}
\label{appendix:pds}

\subsection{Linear Program}

In our setting, a linear program (LP) of $n$ variables and $m$ constraints is an optimization problem where one aims to find a real vector $\Tilde{x}$ that solves the following:

\begin{equation*}\tag{I}\label{eq:LP1}
\begin{array}{ll}
\text{minimize}  & \displaystyle \Tilde{c}^T\Tilde{x}\\
\text{subject to}& \displaystyle \Tilde{A}\Tilde{x} \leq \Tilde{b}\\
                 & \Tilde{l} \leq \Tilde{x} \leq \Tilde{u}
\end{array}
\end{equation*}
where $\Tilde{c}, \Tilde{x}, \Tilde{l}, \Tilde{u}$ are vectors of size $n$, $\Tilde{b}$ is vector of size $m$, and $\Tilde{A}$ is a matrix of size $m \times n$.

For our package query application, we assume that $\Tilde{x}$ is finitely bounded between $\Tilde{l}$ and $\Tilde{u}$. This assumption allows us to finitely bound $\Tilde{b_l} \leq \Tilde{A}\Tilde{x} \leq \Tilde{b_u}$. Rewriting the above LP \eqref{eq:LP1} into a standard form by adding the slack variables represented as a vector $s$ of size $m$, we have:

\begin{equation*}\tag{II}\label{eq:LP2}
\begin{array}{ll}
\text{minimize}  & \displaystyle \Tilde{c}^T\Tilde{x}\\
\text{subject to}& \displaystyle -\Tilde{A}\Tilde{x} + Is = 0\\ 
                 & \Tilde{l} \leq \Tilde{x} \leq \Tilde{u} \\
                 & \Tilde{b_l} \leq s \leq \Tilde{b_u}
\end{array}
\end{equation*}
where $I$ is the identity matrix of size $m \times m$.

Now we convert our original LP of $n$ variables to an LP of $n+m$ variables by concatenating $\Tilde{x}$ and $s$ into $x$.

\begin{equation*}\tag{III}\label{eq:LP3}
\begin{array}{ll}
\text{minimize}  & \displaystyle c^Tx \\
\text{subject to}& \displaystyle Ax = 0\\
                 & l \leq x \leq u
\end{array}
\end{equation*}
where $c=[\Tilde{c} | 0]$, $A=[-\Tilde{A} | I]$, $x=[\Tilde{x} | s]$, $l = [\Tilde{l} | \Tilde{b_l}]$, $u = [\Tilde{u} | \Tilde{b_u}]$.
The LP formulation in \eqref{eq:LP3} is in standard form, with $n+m$ variables and $m$ constraints that our simplex method will solve.

\subsection{Simplex Algorithms}

Since its inception, the simplex algorithm and its variants have dominated the approaches to solving LP problems \cite{Simplex}. Alternative approaches such as the interior-point methods \cite{Interior} have also gained popularity due to better theoretical guarantees. However, in terms of simplicity and ease of parallelization, the simplex algorithm is far more preferable. We focus on the ``bounded'' simplex algorithm for LPs with upper and lower bounds on the components of $x$, as given in \eqref{eq:LP3} above.

Intuitively, the set of feasible solutions to the LP forms a convex \emph{polytope}. (A polytope generalizes a 2D polygon and a 3D polyhedron to $n>3$ dimensions.) A \emph{basic feasible solution} corresponds to a vertex of the polytope. For such a solution, every variable is either:
\begin{itemize}
    \item \emph{nonbasic} and takes values either at the lower bound or upper bound, or 
    \item \emph{basic} and its value can be determined from the nonbasic variables; the set of basic variables is always of size $m$.
\end{itemize}
If there exists an optimal solution to the LP, then at least one basic feasible solution is optimal, and so simplex algorithms search through the basic feasible solutions. For a basic feasible solution, the set $B$, called the \emph{basis}, comprises the indices of the basic variables; its complementary set $N$ is the set of indices of nonbasic variables. We denote by $x_B$ the sub-vector of $x$ containing values of basic variables and by $x_N$ the sub-vector of $x$ containing values of nonbasic variables and similarly define $c_B$ and $c_N$.

It is not practically feasible to examine, in a brute-force manner, all $\binom{n+m}{m}$ basic feasible solutions that correspond to all possible choices for the set of basic variables. This is especially true because, for each combination, we also need to decide whether each nonbasic variable should be set to its lower or upper bound. Instead, the most basic type of simplex method, a ``primal'' simplex method, starts with an initial basis and, for each simplex iteration, swaps a basic variable with a nonbasic one so as to maximally decrease the objective value. The simplex algorithm terminates when no further improvement is possible, and the resulting basic feasible solution is guaranteed to be optimal. Geometrically, this procedure corresponds to starting at some random vertex and then repeatedly moving from the current vertex to an adjacent vertex until an optimal solution is found. (The red arrow on the left side of Figure~\ref{fig:longstep} shows a single step in the case of a 3D polytope.)

The following result gives the necessary and sufficient conditions for a basic feasible solution to be optimal.
\begin{theorem}[Strong duality theorem]
A basic feasible solution with basis $B$  is optimal if and only if $B$ is both primal-feasible and dual-feasible, where
\begin{itemize}
    \item $B$ is primal-feasible if $l_B \leq x_B \leq u_B$
    \item $B$ is dual-feasible if for all nonbasic variables $x_i$ ($i \in N$):
    \begin{itemize}
        \item If $x_i$ is at lower bound then $d_i \geq 0$
        \item If $x_i$ is at upper bound then $d_i \leq 0$
    \end{itemize} where $d=c-(A_B^{-1}A)^Tc_B$ is the reduced cost and $A_B$ is the sub-matrix of $A$ where the columns are indexed by $B$.
\end{itemize}
\end{theorem}

The original simplex algorithm developed by Dantzig \cite{Simplex} is a primal simplex algorithm where it starts with a primal-feasible basis $B$ (which can be dual-feasible or dual-infeasible) and tries to reduce dual infeasibility until the optimality conditions are satisfied. On the other hand, a dual simplex algorithm maintains a dual-feasible basis while reducing primal infeasibility.

\begin{figure}
    \centering
    \includegraphics[width=0.5\textwidth]{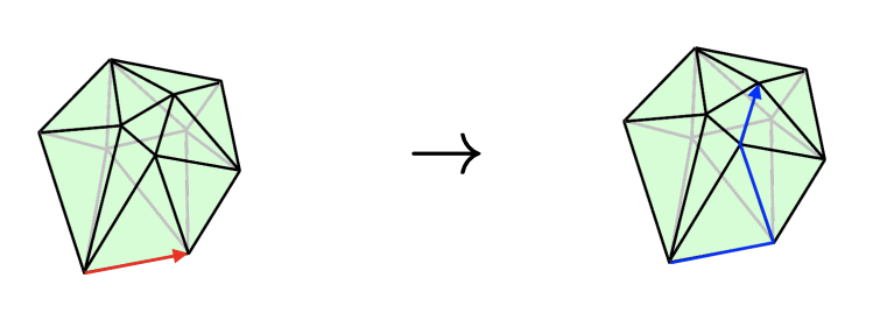}
    \caption{An example of a short-step (left) and a long-step (right) iteration over the same polyhedron.}
    \label{fig:longstep}
\end{figure}

The argument favoring the dual version of simplex over the primal in our setting is the existence of the Bound Flipping Ratio Test (BFRT) procedure. Intuitively, BFRT works well when the geometry of the LP problem allows ``long-step'' iterations. 
As discussed above, the primal simplex algorithm moves from the current vertex to the most beneficial adjacent vertex at each step.

In the dual simplex algorithm with BFRT, an iteration can be equivalent to many such steps and is called a long step. (The long step must be dual-feasible.)
Therefore, the dual version requires many fewer simplex iterations compared to the primal version.

\end{toappendix}


\begin{toappendix}
\section{Parallel dual simplex}\label{appendix:pdsTricks} Given the foregoing considerations, dual simplex is the best choice in our setting since we have found that in many of our LP runs, the geometry of our LP problem enables many such long-step iterations to happen; for example, we usually observe that the first simplex iteration is always a long-step iteration that is equivalent to approximately $n/2$ single-step iterations. 

Below is a discussion of simplifications and optimizations in our implementation of \pds. We repeatedly take advantage of the fact that $m$ is small (on the order of 10--20 constraints) and $n\gg m$.

\subsection{Finding an Initial Dual-Feasible Basis} Recall that the dual simplex algorithm assumes that we start with a dual-feasible basis. Therefore, finding a dual-feasible basis to start with is the main goal of phase-1 of the simplex algorithm. The standard approach to phase-1 is usually a modification to the objective function that penalizes the dual infeasibility of the basis. Hence minimizing such artificial objective function allows us to eventually find a dual-feasible basis. Therefore, phase-1 simplex usually requires the execution of the simplex algorithm itself.

We show that in our setting, phase-1 does not require such execution since the basis is readily available.
Indeed, let us pick the basis as the set of slack variables represented by vector $s$. Then $c_B=0$ since the objective coefficients of the slack variable in our standard form is zero. Hence, the reduced cost $d=c-(A_B^{-1}A)^Tc_B=c$. Hence, to make $B$ dual-feasible, we set
\begin{itemize}
    \item $x_i=l_i$ when $c_i \geq 0$ for $i \in N$.
    \item $x_i=u_i$ when $c_i \leq 0$ for $i \in N$.
\end{itemize}

This procedure gives us $x_N$ which will determine $x_B$ via $$x_B=-A_B^{-1}A_Nx_N$$.

Notice that such a basis is readily available since we have assumed that $\Tilde{x}$ is finitely bounded in the first place.

\subsection{Inverting the Matrix $A_B$} In the dual simplex algorithm, there are important procedures called \textit{FTran} and \textit{BTran} where we have to solve a system of linear equations iof the form $A_Bu=v$ where $u$, $v$ are vectors of size $m$. Standard approaches maintain LU-factorization that stores $A_B=LU$ where $L$ and $U$ are lower and upper triangular matrices, respectively. Whenever a swapping occurs, an update is made to the LU-factorization of $A_B$. In our setting, since the number of constraints $m$ is from 1 to 20 at most, $A_B^{-1}$ can be updated and stored directly. This allows lower memory usage as well as efficient updates.

\subsection{Parallelization Opportunities} In our setting, there are two procedures of the dual simplex that take most of the execution time: \textit{Pivot Computation} (45\%) and \textit{Bound Flipping Ratio Test} (35\%). 

The first procedure involves a matrix-vector multiplication $A_N^Tu$ where $A_N$ is a matrix of size $m \times n$ and $u$ is a vector of size $m$. In our setting, since $n$ can be at a scale of hundreds of millions, it is efficient to parallelize the computation with respect to $n$. 

For the second procedure, the nature of BFRT is sequential. However, in our setting, it is possible to efficiently parallelize it. In layman's terms, BFRT is equivalent to the following problem:

\paragraph{Bill is an enthusiastic traveler who wants to go to as many locations as he can out of $n$ possible locations. Each location $i$ has:}
\begin{itemize}
    \item A scenic score \textbf{si}.
    \item A cost \textbf{ci}.
\end{itemize}
\paragraph{Bill must travel to these locations in ascending order of scenic score without skipping any location while staying within his total budget \textbf{G}. How can we help Bill quickly identify the sequence of locations that he will travel to, given a budget \textbf{G}?} \mbox{}\\

Due to the geometry of our LP problems, in the first simplex iteration, Bill will have a very large budget \textbf{G} which allows him to visit approximately 50\% of the $n$ possible locations. For subsequent iterations, \textbf{G} is much smaller and he can only visit a very small number of locations (0-200). Therefore, we have two parallelization algorithms for large budget and small budget \textbf{G} respectively given $p$ the number of cores used.

\begin{alg} BFRT First Iteration (Large budget \textbf{G})
\begin{itemize}
    \item Compute \textbf{si} and \textbf{ci} for all $n$ locations in parallel with time complexity $O(n/p)$.
    \item Parallel sort the entire set of locations based on \textbf{si} using MapSort by EdaHiro \cite{mapsort} with complexity $O(p^{-1}n\log n)$.
    \item Compute the consecutive sum of sorted \textbf{ci} in parallel with complexity $O\bigl(p+2np^{-1}\bigr)$.
    \item Use binary search on the consecutive sum of \textbf{ci} to find the last location that Bill can go in $O(\log n)$.
\end{itemize}
\end{alg}

Hence the overall time complexity of $O(p^{-1}n\log n)$.

\begin{alg} BFRT Subsequent Iterations (Small budget \textbf{G})
\begin{itemize}
    \item While computing \textbf{si} and \textbf{ci} for all $n$ locations in parallel, each core maintains a max-heap to keep $z$ locations ($z \ll n$ and $z$ can vary among the cores) of the lowest-scoring scenic locations while keeping the sum of \textbf{ci} of these $z$ locations above \textbf{G}. The complexity can be done in $O(p^{-1}n \log z)$.
    \item After computing \textbf{si} and \textbf{ci}, we would have a set of candidate locations collected from all the max-heaps. The size of such a set would approximately be $\Tilde{z}p$ where $\Tilde{z}$ is the average size of the max-heap in each core by the end of the computation.
    \item Use a min-heap to sequentially pop the lowest-scoring scenic locations from the set while keeping the budget within \textbf{G}. This can be done in $O(p \Tilde{z} \log {p \Tilde{z}})$.
\end{itemize}
\end{alg}

Hence the overall time complexity is $O(p^{-1}n \log z)$ where $z$ is the size of the largest max-heap in a core. However, in our observations, such $z$ does not deviate too much from the average $\Tilde{z}$.

\section{\dlv}
\label{appendix:dlv}

\subsection{\textsf{GetScaleFactors} function}
\label{appendix:dlv_scale}

Algorithm \ref{alg:gsf} approximates the constants $c_j$ for each attribute $j$ of a relation $S$ such that \aonedlv partitions a set $P$ over attribute $j$ into approximately $\dfv$ subsets via a bounding variance $\beta=c_j \sigma^2_j / \dfv^2$ where $\sigma^2_j$ is the variance of the attribute $j$ of the tuples in $P$. 

The algorithm first samples a subset $P$ of the relation $S$ (line~1). For each attribute $j$, it computes the minimum and maximum possible variance of a subset of $P$ (line~3,4) and then runs a binary search of the bounding variance $\beta$ on this interval to compute $\beta$ such that \aonedlv partitions $P$ into approximately $n \approx \dfv$ subsets (line~7). 

\begin{algorithm}[H]
\caption{\textsf{GetScaleFactors}}\label{alg:gsf}
\begin{algorithmic}[1]

\Require $S := $ set of $k$-dimensional tuples
\Statex $\dfv := $ \df

\Ensure $N := $ sampling size
\Statex $\epsilon := $ epsilon for binary search

\State Uniformly sample $P \subseteq S$ such that $|P|=\min(|N|, |S|)$
\For{each attribute $j \in [1,\ldots,k]$}
    \State $\sigma^2_{\text{min}} \gets 0$
    \State $\sigma^2_{\text{max}} \gets \frac{1}{4}(\max_{t \in P}t.j - \min_{t \in P} t.j)^2$
    \While{$|\sigma^2_{\text{max}}-\sigma^2_{\text{min}}| > \epsilon$} \Comment{ {\color{gray}\textit{Binary search on $\beta$}}}
        \State $\beta \gets (\sigma^2_{\text{max}}+\sigma^2_{\text{min}})/2$
        \State $n \gets |\onedlv(\beta,P,j)|$
        \If{$n=\dfv$}
            \State \textbf{break}
        \ElsIf{$n<\dfv$}
            \State $\sigma^2_{\text{max}} \gets \beta$
        \Else
            \State $\sigma^2_{\text{min}} \gets \beta$
        \EndIf
    \EndWhile
    \State $\sigma_j^2 \gets$ the variance of the attribute $j$ of the tuples in $P$
    \State $c_j \gets \beta \dfv^2 / \sigma_j^2$
\EndFor

\Return $(c_1,c_2,\ldots,c_k)$

\end{algorithmic}
\end{algorithm}

\subsection{\dlv in Large Relations}
\label{appendix:dlv_large}

When designing algorithms for large relations, three important desiderata are: (1)~ability to run on limited in-memory storage, (2)~cache-friendliness, and (3)~high degree of parallelization. In \lsr, it is required to produce a $p$-partition where $p \approx n/\dfv$. Algorithm~\ref{alg:dlv} only satisfies requirement~(2) and cannot run on limited in-memory storage since it needs to store at least $p$ sets of partitioning information in a max priority queue $\xP$. We can extend \adlv to satisfy the three desiderata via a bucketing scheme~\cite{bucket}. Specifically, supposing that $r$ tuples can fit into memory, we choose an attribute with the highest variance and partition its range into a number of equally spaced buckets such that each bucket has at most $r$ tuples. Then for each bucket, we can reliably execute Algorithm~\ref{alg:dlv} as usual. 

The extended \adlv algorithm is still cache-friendly since the majority of the work is via Algorithm~\ref{alg:dlv}. It can be run on limited in-memory storage by choosing an appropriate value of $r$. Because bucketing is highly parallelizable, we only have to parallelize Algorithm~\ref{alg:dlv}. The simplest solution is to provide a locking mechanism for modifying the max-priority queue $\xP$.

\neighbor also requires fast group membership queries for arbitrary tuples. Specifically, each partitioning subset---i.e., each group in the relation---needs to store information about the ranges of each of its attributes to quickly determine if a tuple belongs within the group.

Our \adlv implementation stores partitioning information such as the mean, the variance, and the range of each attribute in a \pg table~\cite{postgres}. Ranges are represented using \pg's built-in range data type to enable the efficient execution of range queries, i.e., determining if a point is contained in an interval. \pg also has a Generalized Search Tree (GiST) index~\cite{gist} which further accelerates these range queries, achieving sub-linear running time. Our \adlv creates a multi-column GiST index on all the columns in order of highest variance column first. A membership query on the GiST index of a table of size 10 million runs in 0.5ms on average. This implementation choice allows our \adlv to be stored compactly without needing a full representation of a tree-like structure as with \kd. 

\section{Handling Local Predicates}
\label{appendix:local}

Local predicates are constraints that each tuple in the package has to satisfy individually. Depending on the selectivity of the local predicates, the representative tuples in \adlv can become increasingly inaccurate. For example, if relatively few tuples satisfy the local predicates (low selectivity), then each representative tuple might represent fewer tuples than it should, so its attribute values may not accurately represent the few tuples that satisfy the predicates.

When the workload of package-query users involves a fixed set of local predicates or when the local predicates have low selectivity leading to inaccuracy, it may be beneficial to do \adlv partitioning after applying the local predicates, thereby avoiding the risk of inaccurate representative tuples. However, if the local predicates change from query to query, running \adlv partitioning for every new set of local predicates is costly. We, therefore, propose two possible strategies---to be analyzed in future work---for handling local predicates efficiently.

A naive approach runs \adlv once over all tuples in the relation and then, given a set of local predicates, simply adjusts the representative tuples throughout the hierarchy of relations.
For example, starting from layer~0 (the original relation), we select only tuples that satisfy the local predicates and
recompute each representative tuple as the mean of the selected tuples in its represented group; if there are no tuples selected within a group~$l$, there will be no representative tuple for that group in layer~$l+1$. This approach, in the worst case, requires us to materialize the hierarchy of relations but it does avoid the risk of inaccurate representative tuples.

A more efficient but less accurate approach executes \lsr as if there were no local predicates while going from layer~$L$ to layer~1. Then, when going from layer~1 to layer~0 using \shading (Algorithm~\ref{alg:shading}), the \neighbor algorithm is adjusted by modifying the \textsf{GetTuples}$(1,g)$ function to return only those layer-0 tuples in group~$g$ that satisfy the local predicates.
This extra cost is minimal and we hope that the relatively large set (up to size $\alpha$) of layer-0 tuples would be rich enough to largely compensate for any inaccuracies in the representative tuples in layers~1 through $L$. 

\section{Experiments}
\label{appendix:exp}


\subsection{Additional queries} 
\label{appendix:ae}

\begin{figure}[tb]
\centering
\includegraphics[width=\columnwidth]{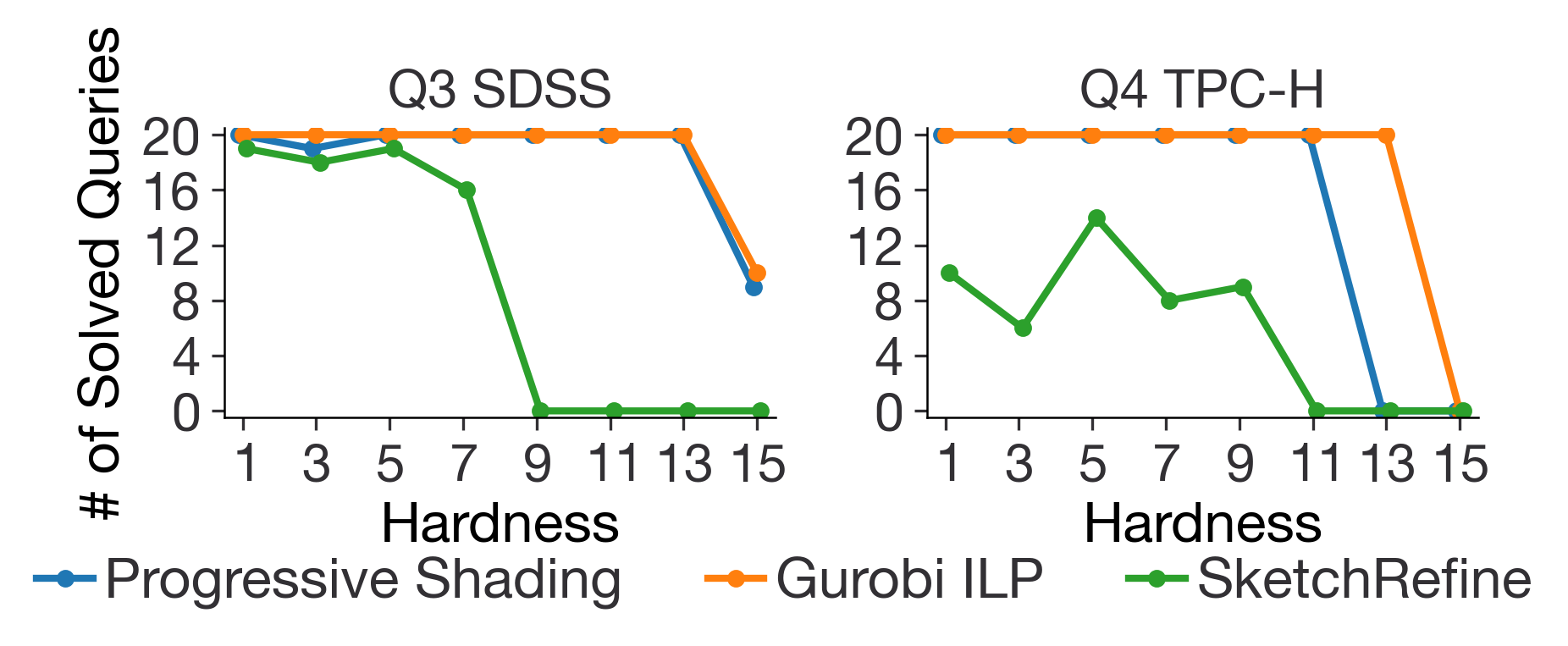}
\caption{False infeasibility as hardness increases for Q3 \ssds and Q4 \tpch.}\label{fig:e2}
\end{figure}

We present two additional query structures called Q3 \ssds and Q4 \tpch in Table~\ref{tab:benchmark2}. We show the query performance as relation size increases in Figure~\ref{fig:e1} and false infeasibility as hardness increases in Figure~\ref{fig:e2}. 

In both additional queries, \gb only scales up to one million tuples, and the running time grows exponentially with the relation size. \sr falls out of scale beyond 10 million but manages to find solutions in all hardnesses even though they are sub-optimal in both queries. In contrast, \lsr always finds solutions with a low integrality gap with high probability. \sr's solutions in Q4 \tpch have abnormally high integrality gaps because the LP solutions have an objective value of 0: the integrality gap of any sub-optimal ILP solution found by \sr is scaled up by a factor of $1/\epsilon=10$. 

In query Q4 of \tpch at hardness 3 (relation size = $10^8$) and at hardness 7 (relation size = $10^9$), we observe a rare phenomenon where \lsr with 80 cores produces sub-optimal results (the upper IQR values of the integrality gap are $10^4$) compared to \lsr with 4 cores for at most 3 runs. This happens because of race conditions within the loop and in lines 11-13 of \neighbor (Algorithm \ref{alg:neighbor_group}): faster cores can add worse objective value tuples into $S'_l$ (line 11), which are then returned before slower cores get to add better objective value tuples. If the \neighbor misses a few crucial tuples with better objective values, the overall solution can be sub-optimal with a high integrality gap when compared to the zero objective value LP solution of Q4 \tpch.

For Figure~\ref{fig:e2}, as hardness increases, false infeasibility is observed in both \lsr and \sr. For \sr, solutions are hardly found when $\hard>7$ in both queries while \lsr only struggles with $\hard=13$ in Q4 \tpch. 

\begin{figure}[htb]
\centering
\includegraphics[width=\columnwidth]{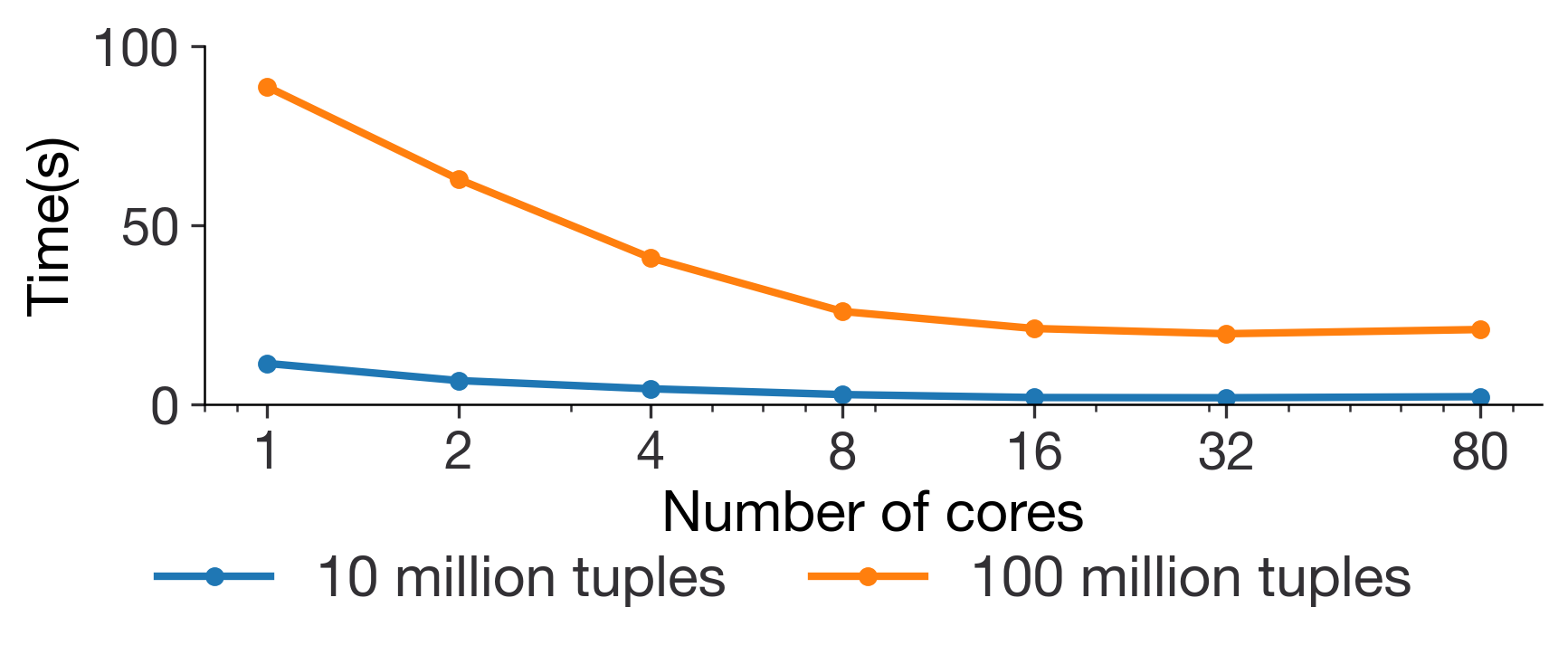}
\caption{Mini-Experiment 3. \textit{How well does \pds scale with more cores?} The performance of \pds scales relatively well up to 16 cores. Nevertheless, at 80 cores, \pds achieves 4.79x speed up.}
\label{fig:m3}
\end{figure}

\begin{table*}[t]
\begin{adjustbox}{max width=\textwidth, center}
\begin{tabular}[t]{@{}ccc|ccccc|ccc|ccccc@{}}
\toprule
\multicolumn{8}{c|}{Q3 \ssds}
& \multicolumn{8}{c}{Q4 \tpch}
\\ \midrule

\multicolumn{8}{l|}{
{\small
\begin{tabular}[t]{@{}l@{}}
{\tt SELECT  \textbf{PACKAGE}(*) AS P FROM sdss R \textbf{REPEAT} 0}\\ 
{\tt \textbf{SUCH THAT}   25 $\leq$ COUNT(P.*) $\leq$ 75 AND}\\  
{\tt SUM(P.tmass\_prox) $\geq b_1$ AND SUM(P.j) $\leq b_2$ AND}\\     
{\tt SUM(P.h) BETWEEN $b_3$ AND $b_4$} \\ 
{\tt \textbf{MAXIMIZE} SUM(P.k)}
\end{tabular}}
} & 
\multicolumn{8}{l}{
{\small
\begin{tabular}[t]{@{}l@{}}
{\tt SELECT \textbf{PACKAGE}(*) AS P FROM tpch R \textbf{REPEAT} 0}\\ 
{\tt \textbf{SUCH THAT}   50 $\leq$ COUNT(P.*) $\leq$ 150 AND}\\   
{\tt SUM(P.quantity) $\leq b_1$ AND}\\ 
{\tt SUM(P.price) BETWEEN $b_2$ AND $b_3$}\\ 
{\tt \textbf{MINIMIZE} SUM(P.tax)}
\end{tabular}}
} \\

\midrule

\textbf{Attribute} & $\mu$ & $\sigma$ & $\hard$: & 1 & 3 & 5 & 7 &
\textbf{Attribute} & $\mu$ & $\sigma$ & $\hard$:  & 1 & 3 & 5 & 7  \\

{\tt k} & 13.73 & 1.727 &  &  &  &  &  & 
{\tt tax} & 1530 & 1485 &  &  &  &  & \\

{\tt tmass$\_$prox} & 14.45 & 14.96 & $b_1$ & 732.02 & 858.07 & 936.48 & 997.69 & 
{\tt quantity} & 25.50 & 14.43 & $b_1$ & 2480.985 & 2281.968 & 2155.994 & 2056.884  \\

{\tt j} & 14.82 & 1.562 & $b_2$ & 740.01 & 726.85 & 718.66 & 712.27 & 
{\tt price} & 38240 & 23290 & $b_2$ & 3729135 & 3814767 & 3823077 & 3823908  \\

{\tt h} & 14.05 & 1.657 & $b_3$ & 695.25 & 701.03 & 702.18 & 702.43 & 
 &  &  & $b_3$ & 3918865 & 3833233 & 3824923 & 3824092  \\
 
 &  &  & $b_4$ & 709.75 & 703.97 & 702.82 & 702.57 &  &  &  
 &  & &  & &  \\

\bottomrule
\end{tabular}%
\end{adjustbox}
\caption{Experimental Benchmark Q3 \ssds and Q4 \tpch. The package query templates, underlying data statistics, and constraint bounds at different query-hardness ($\hard$) levels.}
\label{tab:benchmark2}
\end{table*}

\smallskip\noindent

\begin{figure*}
    \centering
    \includegraphics[width=1\textwidth]{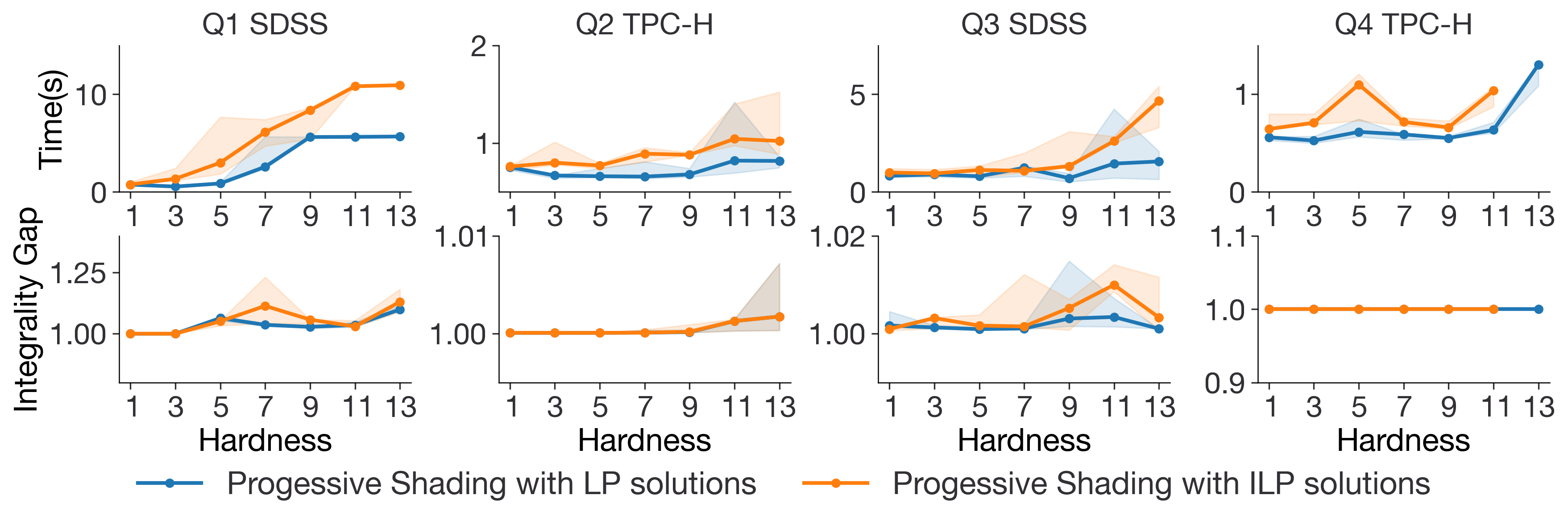}
    \caption{Mini-Experiment 1. \textit{Does replacing the LP solution with an ILP solution in \shading improve overall optimality?} There is no difference in terms of the optimality or solvability of \lsr with LP solution or ILP solution. We choose \lsr with LP solutions since they are faster in all cases.}
    \label{fig:m1}
\end{figure*}

\subsection{Mini-Experiments}
\label{appendix:mini}

The results of Mini-Experiments~1, 2, 3, and 4 in the main text are plotted in Figures~\ref{fig:m1}, Figures~\ref{fig:m2}, Figures~\ref{fig:m3}, and Figures~\ref{fig:m4} respectively. 

\begin{expbox}
\noindent\textbf{Mini-Experiment 6.} \textit{What is the optimal configuration of \protect\aug \protect\augv and \protect\df \protect\dfv?} 

We explored the performance of \lsr and \dlv over 9 configurations of \protect\aug \protect\augv and \protect\df \protect\dfv (Table~\ref{tab:grid}). We ran the queries from Q1 \ssds and Q2 \tpch described in Table~\ref{tab:benchmark} with query-hardness levels $\hard \in \{1,3,5,7\}$ (Section~\ref{sec:Esetup}). For each $\hard$, we randomly sampled 5 sub-relations of size 10 million representing 5 queries for a total of 20 queries per query per configuration. As shown in Table~\ref{tab:grid}, $\augv=100{,}000$ and $\dfv=100$ is the optimal configuration.
\end{expbox}


\begin{expbox}
\noindent\textbf{Mini-Experiment 7.} \textit{What is the optimal sub-ILP size $q$ for \filp?} 

We explored the performance of \filp as we varied the sub-ILP size $q$ in $\{50, 500, 5000, 50000\}$ (Figure~\ref{fig:q}). We ran all the 4 queries with query-hardness levels $\hard \in \{1,3,5,7,9,11,13\}$ (Section~\ref{sec:Esetup}). For each \hard, we randomly sampled 10 sub-relations of size $100$ thousand representing 10 queries for a total of 70 queries per query per value of $q$. As shown in Figure~\ref{fig:q}, $q=500$ is the optimal sub-ILP size for \filp. 

\end{expbox}

\begin{expbox}
\textbf{Mini-Experiment 8. }\textit{Is it worth using \protect\filp for the last layer of \protect\lsr instead of \protect\gb?} 

We explored the performance of \lsr with \filp and with \gb (Figure~\ref{fig:m6}). We ran all the 4 queries with query-hardness levels $\hard \in \{1,3,5,7,9,11,13\}$ (Section~\ref{sec:Esetup}). For each \hard, we randomly sampled 5 sub-relations of size 100 million representing 5 queries for a total of 35 queries per query. As shown in Figure~\ref{fig:m6}, we conclude that using \filp significantly improves the time execution of \lsr when the query hardness is above 7 while maintaining relatively low integrality gaps.
\end{expbox}

\begin{figure*}
    \centering
    \includegraphics[width=1\textwidth]{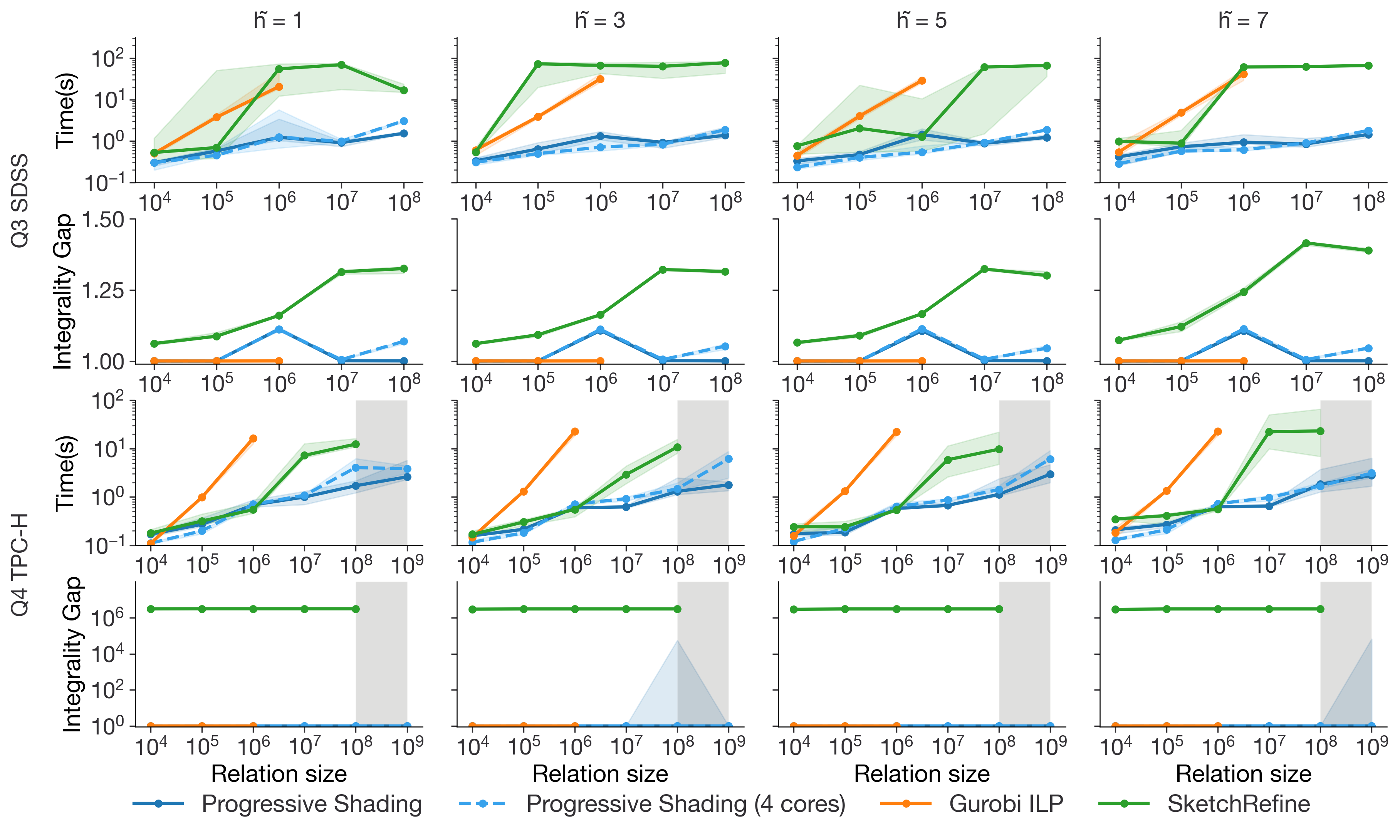}
    \caption{Query performance as relation size increases for Q3 \ssds and Q4 \tpch. In both measurements, the values are plotted as the median of the 10 runs and the error bands are the interquartile range (IQR) of the 10 runs.}
    \label{fig:e1}
\end{figure*}

\begin{figure*}
    \centering
    \includegraphics[width=1\textwidth]{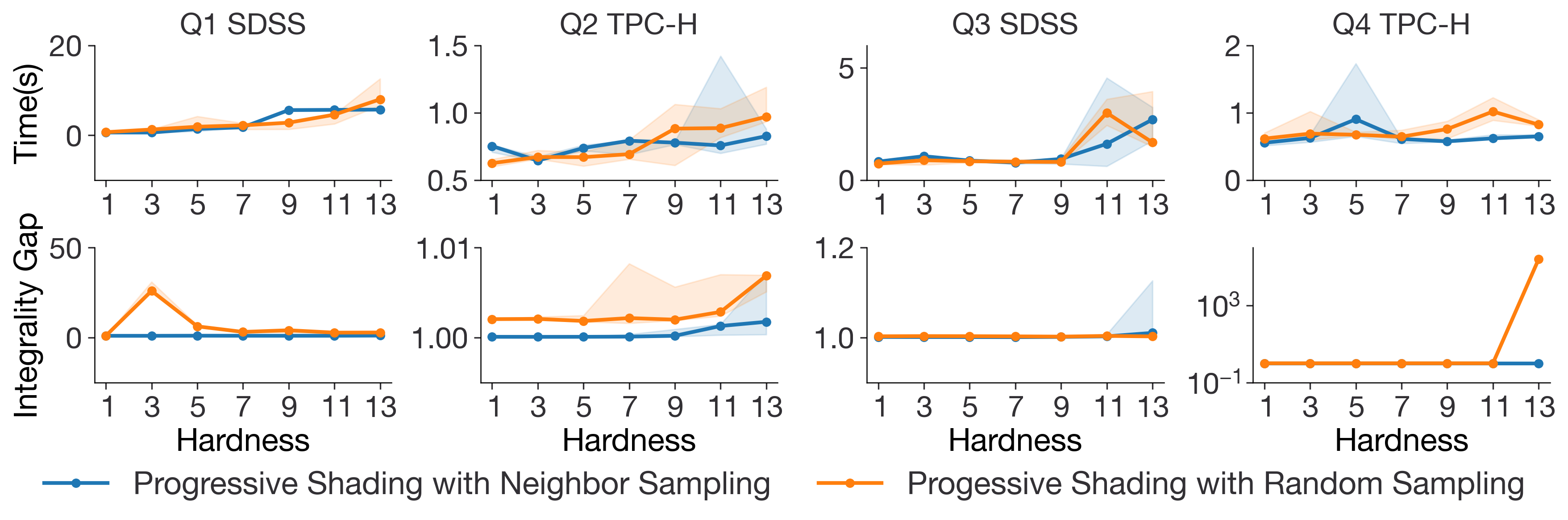}
    \caption{Mini-Experiment 2. \textit{Does replacing \neighbor with a random sampling of representative tuples impact the overall performance of \lsr?} \lsr with \neighbor has much more optimal solutions than \lsr with Random Sampling in Q1 \ssds and Q4 \tpch.}
    \label{fig:m2}
\end{figure*}

\begin{figure*}
    \centering
    \includegraphics[width=1\textwidth]{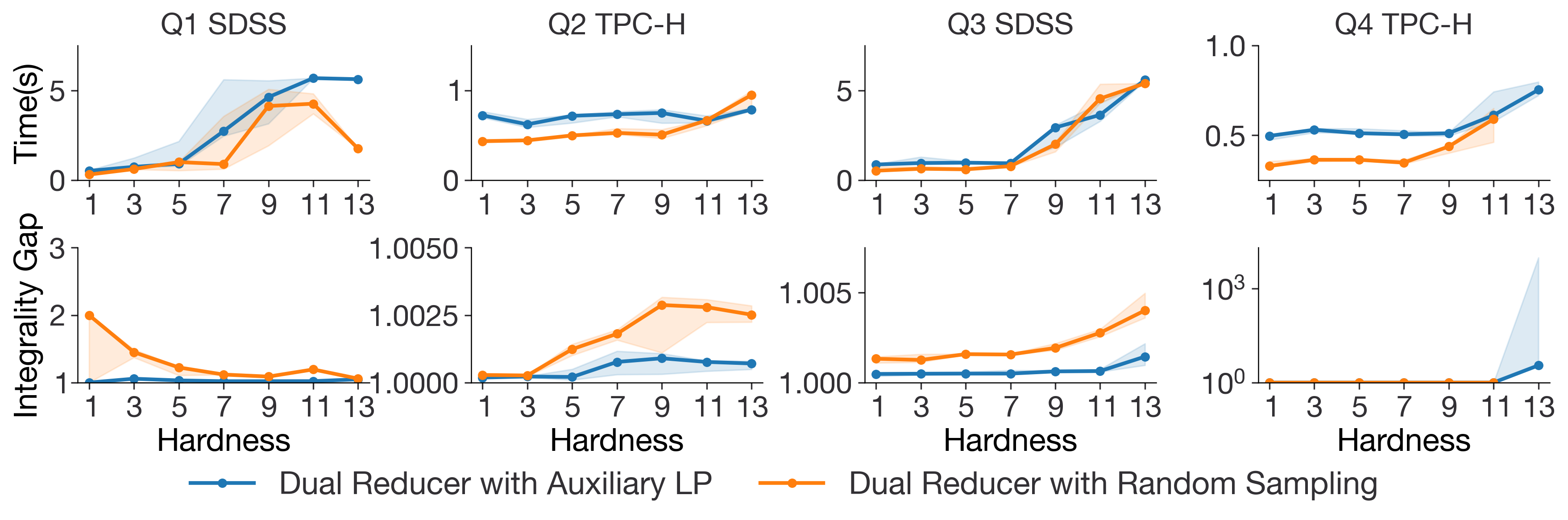}
    \caption{Mini-Experiment 4. \textit{Does replacing the Auxiliary LP with a random sampling of tuples from $S$ to formulate a sub-ILP of size $q$ impact the overall performance of \filp?} \filp with Auxiliary LP showed improved solvability at $\hard=13$ (Q4 \tpch) and provided solutions with better optimality than \filp with random sampling in all queries.}
    \label{fig:m4}
\end{figure*}

\begin{table*}[t]
\begin{adjustbox}{max width=\textwidth, center}
\begin{tabular}{|ccc|cccccc|}
\hline
\multicolumn{3}{|c|}{}                                                                                                                   & \multicolumn{6}{c|}{Downscale Factor $\dfv$}                                                                                                                                                                           \\ \cline{4-9} 
\multicolumn{3}{|c|}{}                                                                                                                   & \multicolumn{2}{c|}{10}                                      & \multicolumn{2}{c|}{100}                                                                                     & \multicolumn{2}{c|}{1000}                \\ \cline{4-9} 
\multicolumn{3}{|c|}{\multirow{-3}{*}{}}                                                                                                 & $Q1 \ssds$            & \multicolumn{1}{c|}{$Q2 \tpch$}            & $Q1 \ssds$                                    & \multicolumn{1}{c|}{$Q2 \tpch$}                                    & $Q1 \ssds$            & $Q2 \tpch$             \\ \hline
\multicolumn{1}{|c|}{}                                           & \multicolumn{1}{c|}{}                         & Query Time (s)        & $1.0415 \pm 0.095$ & \multicolumn{1}{c|}{$1.3595 \pm 0.095$} & $0.57859 \pm 0.17$                         & \multicolumn{1}{c|}{$0.60817 \pm 0.075$}                        & $1.1243 \pm 1.0$   & $0.80364 \pm 0.105$ \\
\multicolumn{1}{|c|}{}                                           & \multicolumn{1}{c|}{}                         & Partitioning Time (s) & $35.266 \pm 0.11$  & \multicolumn{1}{c|}{$33.063 \pm 0.24$}  & $13.762 \pm 0.16$                          & \multicolumn{1}{c|}{$12.96 \pm 0.14$}                           & $14.065 \pm 0.17$  & $13.451 \pm 0.85$   \\
\multicolumn{1}{|c|}{}                                           & \multicolumn{1}{c|}{}                         & Integrality Gap       & $1.0561 \pm 0.115$ & \multicolumn{1}{c|}{$1.0001 \pm 0.0$}   & $3.5665 \pm 2.33$                          & \multicolumn{1}{c|}{$1.0001 \pm 0.0$}                           & $7.728 \pm 9.79$   & $1.0288 \pm 0.015$  \\
\multicolumn{1}{|c|}{}                                           & \multicolumn{1}{c|}{\multirow{-4}{*}{$10^4$}} & Solve Rate            & 20/20              & \multicolumn{1}{c|}{20/20}              & 20/20                                      & \multicolumn{1}{c|}{20/20}                                      & 19/20              & 20/20               \\ \cline{2-9} 
\multicolumn{1}{|c|}{}                                           & \multicolumn{1}{c|}{}                         & Query Time(s)         & $7.6676 \pm 0.735$ & \multicolumn{1}{c|}{$3.6318 \pm 0.245$} & \cellcolor[HTML]{9eff9e}$1.142 \pm 1.68$   & \multicolumn{1}{c|}{\cellcolor[HTML]{9eff9e}$0.7504 \pm 0.025$} & $2.3068 \pm 1.9$   & $0.69396 \pm 0.125$ \\
\multicolumn{1}{|c|}{}                                           & \multicolumn{1}{c|}{}                         & Partitioning Time(s)  & $35.266 \pm 0.11$  & \multicolumn{1}{c|}{$33.063 \pm 0.24$}  & \cellcolor[HTML]{9eff9e}$13.762 \pm 0.16$  & \multicolumn{1}{c|}{\cellcolor[HTML]{9eff9e}$12.96 \pm 0.14$}   & $14.065 \pm 0.17$  & $13.451 \pm 0.85$   \\
\multicolumn{1}{|c|}{}                                           & \multicolumn{1}{c|}{}                         & Integrality Gap       & $1.0 \pm 0.04$     & \multicolumn{1}{c|}{$1.0001 \pm 0.0$}   & \cellcolor[HTML]{9eff9e}$1.1068 \pm 0.245$ & \multicolumn{1}{c|}{\cellcolor[HTML]{9eff9e}$1.0001 \pm 0.0$}   & $3.8156 \pm 2.04$  & $1.0001 \pm 0.0$    \\
\multicolumn{1}{|c|}{}                                           & \multicolumn{1}{c|}{\multirow{-4}{*}{$10^5$}} & Solve Rate            & 19/20              & \multicolumn{1}{c|}{20/20}              & \cellcolor[HTML]{9eff9e}20/20              & \multicolumn{1}{c|}{\cellcolor[HTML]{9eff9e}20/20}              & 20/20              & 20/20               \\ \cline{2-9} 
\multicolumn{1}{|c|}{}                                           & \multicolumn{1}{c|}{}                         & Query Time(s)         & $36.248 \pm 0.98$  & \multicolumn{1}{c|}{$17.201 \pm 0.485$} & $2.6534 \pm 0.79$                          & \multicolumn{1}{c|}{$3.1335 \pm 0.08$}                          & $2.3494 \pm 0.435$ & $2.1545 \pm 0.16$   \\
\multicolumn{1}{|c|}{}                                           & \multicolumn{1}{c|}{}                         & Partitioning Time(s)  & $35.266 \pm 0.11$  & \multicolumn{1}{c|}{$33.063 \pm 0.24$}  & $13.762 \pm 0.16$                          & \multicolumn{1}{c|}{$12.96 \pm 0.14$}                           & $14.065 \pm 0.17$  & $13.451 \pm 0.85$   \\
\multicolumn{1}{|c|}{}                                           & \multicolumn{1}{c|}{}                         & Integrality Gap       & $1.0181 \pm 0.035$ & \multicolumn{1}{c|}{$1.0001 \pm 0.0$}   & $1.0183 \pm 0.05$                          & \multicolumn{1}{c|}{$1.0001 \pm 0.0$}                           & $1.4807 \pm 0.64$  & $1.0001 \pm 0.0$    \\
\multicolumn{1}{|c|}{\multirow{-12}{*}{Augmenting Size $\augv$}} & \multicolumn{1}{c|}{\multirow{-4}{*}{$10^6$}} & Solve Rate            & 19/20              & \multicolumn{1}{c|}{20/20}              & 20/20                                      & \multicolumn{1}{c|}{20/20}                                      & 20/20              & 20/20               \\ \hline
\end{tabular}
\end{adjustbox}
\caption{Mini-Experiment 6. \textit{What is the optimal configuration of \protect\aug \protect\augv and \protect\df \protect\dfv?} We find that $\protect\augv=100{,}000$ and $\protect\dfv=100$ to be optimal (the green-colored configuration). Lower $\protect\dfv$ would cause the partitioning time to be much higher (almost 3x longer) while higher $\protect\dfv$ would cause the partitioning groups to be less accurate since each group would contain more tuples. Higher $\protect\augv$ significantly increases the query time with marginal gains to solution quality. Lower $\protect\augv$ results in a significant drop in optimality (3x worse).} 
\label{tab:grid}
\end{table*}

\begin{figure*}
    \centering
    \includegraphics[width=\textwidth]{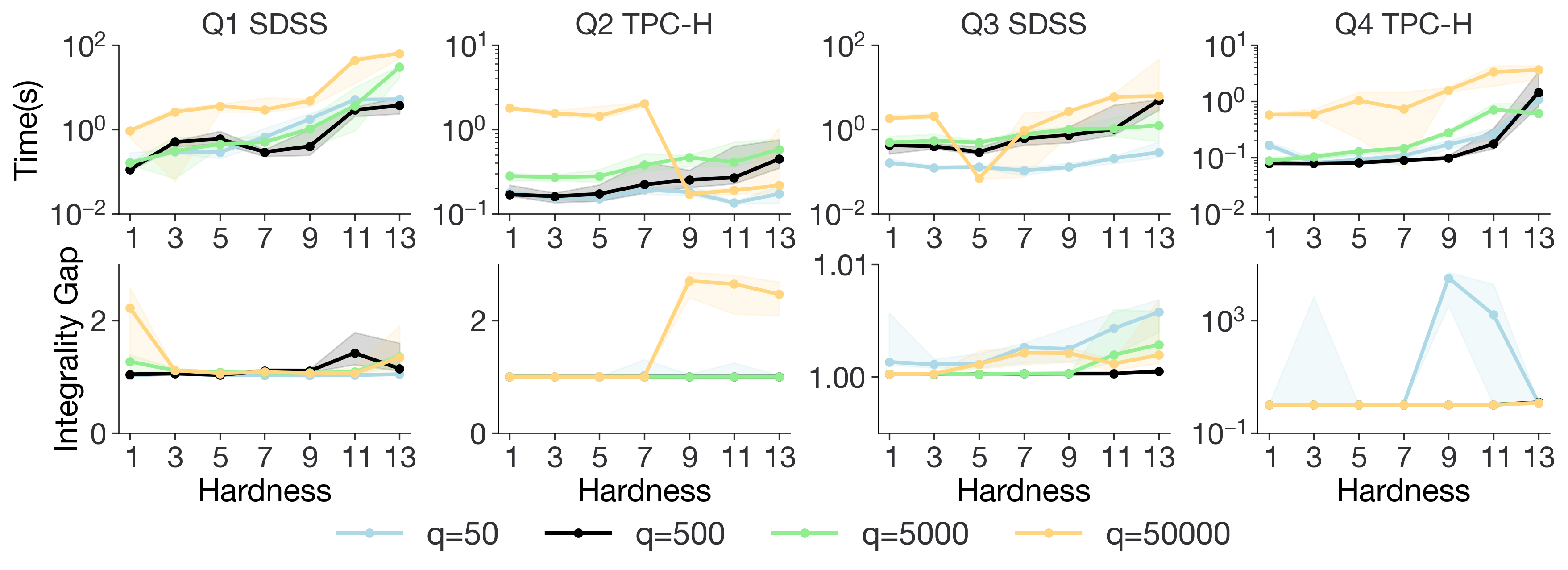}
    \caption{Mini-Experiment 7. \textit{What is the optimal sub-ILP size $q$ for \protect\filp?}. $q=500$ has a faster time execution overall compared to $q=5000$ and $q=50000$. In addition, $q=50$ performs much worse in Q4 \tpch in terms of optimality.}
    \label{fig:q}
\end{figure*}

\begin{figure*}
    \centering
    \includegraphics[width=1\textwidth]{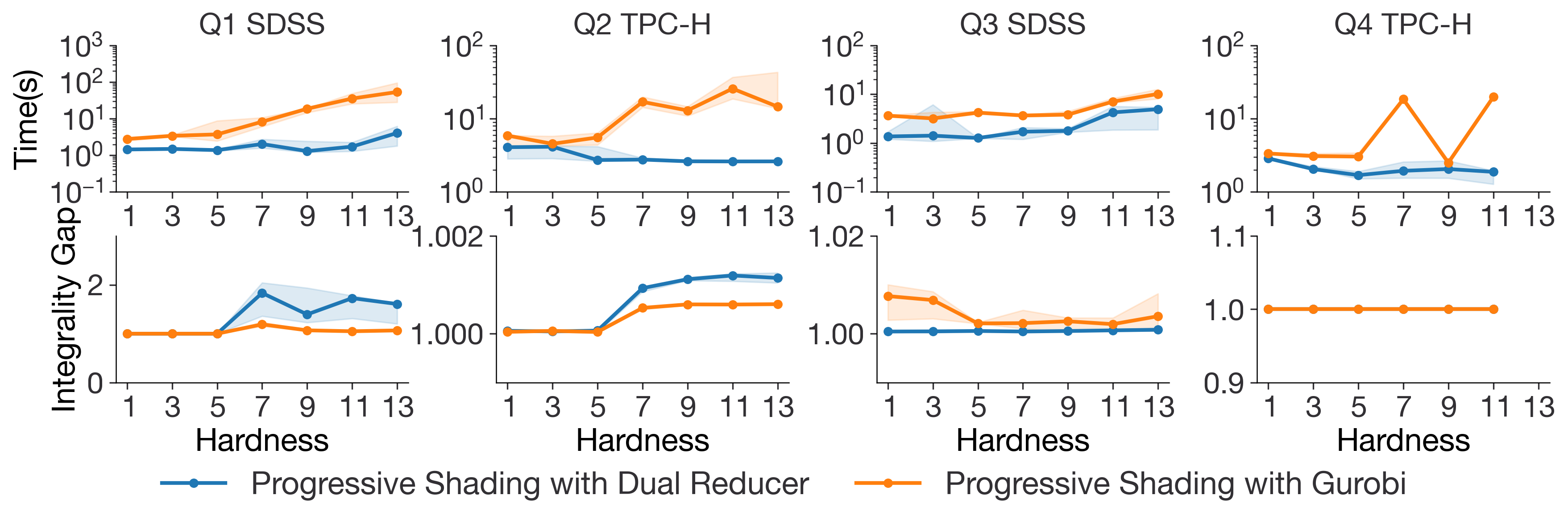}
    \caption{Mini-Experiment 8. \textit{Is it worth using \filp for the last layer of \lsr instead of \gb?} In all queries, \lsr with \filp is much faster than \lsr with \gb when hardness is above 7 while maintaining relatively low integrality gaps.}
    \label{fig:m6}
\end{figure*}

\end{toappendix}

\bibliographystyle{ACM-Reference-Format}

\bibliography{refs}

\newpage
\appendix

\balance

\end{document}